\documentclass[twocolumn]{aa}

\date{Received 18/03/2025 / Accepted 29/05/2025 }

\usepackage[varg]{txfonts}
\usepackage[utf8]{inputenc}
\usepackage[english]{babel}
\usepackage{tabularx}

\usepackage{setspace}

\usepackage{indentfirst}

\usepackage{enumitem}
\setlist[itemize]{noitemsep, topsep=0pt}

\usepackage{graphicx}
\graphicspath{{/Users/elka127/Documents/uio/flat_paper2AA/images}}

\usepackage{wrapfig}
\usepackage{subfigure}
\usepackage[dvipsnames]{xcolor}

\usepackage{hyperref}
\hypersetup{
    colorlinks,
    urlcolor=blue,
    linkcolor=blue,
    citecolor=blue
}

\usepackage{csquotes}

\usepackage{listings}
\usepackage{courier}\definecolor{bluekeywords}{rgb}{0,0,1}
\definecolor{greencomments}{rgb}{0,0.5,0}
\definecolor{redstrings}{rgb}{0.64,0.08,0.08}
\definecolor{types}{rgb}{0.17,0.57,0.68}
\lstdefinestyle{py}
{
    language=Python,
    frame=l,
    framesep=5pt,
    captionpos=b,
    numbers=left,
    numberstyle=\tiny,
    showspaces=false,
    showtabs=false,
    breaklines=true,
    showstringspaces=false,
    breakatwhitespace=true,
    commentstyle=\color{greencomments},
    keywordstyle=\color{bluekeywords},
    stringstyle=\color{redstrings},
    basicstyle=\footnotesize\ttfamily,
}
\newcommand\Tstrut{\rule{0pt}{2.6ex}}       \newcommand\Bstrut{\rule[-0.9ex]{0pt}{0pt}} \newcommand{\TBstrut}{\Tstrut\Bstrut} \usepackage{natbib}
\bibpunct{(}{)}{;}{a}{}{,}
\begin{document}
\title{Flare frequency in M dwarfs belonging to Young Moving Groups.}
\author{E. Mamonova\inst{1}, Y. Shan \inst{1},  A. F. Kowalski\inst{2,3,4}, S. Wedemeyer\inst{5,6}, S. C. Werner\inst{1}}

\institute{Centre for Planetary Habitability (PHAB), University of Oslo, 0315 Oslo, Norway
\and National Solar Observatory, University of Colorado Boulder, 3665 Discovery Drive, Boulder, CO 80303, USA
\and Department of Astrophysical and Planetary Sciences, University of Colorado, Boulder, 2000 Colorado Ave, CO 80305, USA
\and Laboratory for Atmospheric and Space Physics, University of Colorado Boulder, 3665 Discovery Drive, Boulder, CO 80303, USA
\and Rosseland Centre for Solar Physics, University of Oslo, 0315 Oslo, Norway
\and Institute of Theoretical Astrophysics, University of Oslo, 0315 Oslo, Norway}

\abstract{\textit{Context.} M stars are preferred targets for studying terrestrial exoplanets, for which we hope to obtain their atmosphere spectra in the next decade. However, M dwarfs have long been known for strong magnetic activity and the ability to frequently produce optical, broadband emission flares.\\
\textit{Aims.} We aim to characterise the flaring behaviour of young M dwarfs in the temporal, spectral, and energetic dimensions, as well as examine the stellar parameters governing this behaviour, in order to improve our understanding of the energy and frequency of the flare events capable of shaping the exoplanet atmosphere.\\
\textit{Methods.} Young Moving Group (YMG) members provide a unique age-based perspective on stellar activity. By examining their flare behaviour in conjunction with rotation, mass, and H$\alpha$ data, we obtain a comprehensive understanding of flare activity drivers in young stars.\\
\textit{Results.} We demonstrate that young stars sharing similar stellar parameters can exhibit a variety in flare frequency distributions and that the flare behaviour shows indications of difference between optical and far-UV. We propose that the period of rotation, not the age of the star, can be a good proxy for assessing flaring activity. Furthermore, we recommend that instead of a simple power law for describing the flare frequency distribution, a piecewise power law be used to describe mid-size and large flare distributions in young and active M dwarfs.\\
\textit{Conclusions.} Using known periods of rotation and fine-tuned power laws governing the flare frequency, we can produce a realistic sequence of flare events to study whether the atmosphere of small exoplanets orbiting M dwarf shall withstand such activity until life can emerge.}
\keywords{stars: flare -- stars: pre-main sequence  --  stars: low-mass –- planets and satellites: atmospheres -- methods: numerical}
\titlerunning{FFDs in YMG low-mass stars}
\authorrunning{Mamonova et al.}
\maketitle

\section{Introduction}
\label{sec:introduction}

M dwarfs, the most abundant stars in our galaxy, have become focal points in the search for habitable exoplanets \citep{2013ApJ...762...41G, 2015ApJ...807...45D}. These low-mass stars are particularly intriguing for exoplanet searches due to their close-in habitable zones, which facilitate the detection of potentially habitable worlds \citep{2013A&A...556A.110B}. However, M dwarfs are also characterised by high levels of magnetic activity, a factor that significantly impacts planetary habitability \citep{2007AsBio...7..185L,2010AsBio..10..751S,2019AsBio..19...64T}.

Stellar flares, sudden releases of magnetic energy resulting in increased emission across the electromagnetic spectrum, are one of the most prominent manifestations of stellar magnetic activity \citep{1996SSRv...75..453N}. On M dwarfs, these flares can be particularly energetic relative to the star's quiescent luminosity, potentially having profound effects on the atmospheres and surface conditions of orbiting planets \citep{2017ApJ...843..110R}. Understanding the frequency and energy distribution of flares on M dwarfs is therefore crucial for modelling the chemical impact on atmospheres for planets in such systems, which could affect the planet's evolution and, in the end, this would help to asses the habitability of the planetary systems and the potential for life to emerge and persist around red dwarfs \citep{2016ApJ...830...77V}.

Young Moving Groups (YMGs) provide an excellent laboratory for studying the evolution of stellar activity \citep{1997A&A...327.1039C}. These groups of stars, sharing common space motions and ages, allow us to probe how magnetic activity changes as stars evolve within several hundred million years after their birth \citep{2021ApJ...911..111P}. M dwarfs in YMGs are particularly valuable targets, as they represent the early stages of stellar evolution when magnetic activity is expected to be at its peak \citep{2014PASP..126..398H}. By studying these young systems, we can gain insights into the early magnetic evolution of M dwarfs and its implications for planetary habitability during the crucial early phases of planetary system formation and evolution.

In this study, we explore the relationship between stellar rotation period, mass, age, and flaring activity for these young M dwarfs. This period-mass-age relation is crucial for understanding the evolution of stellar magnetic activity, as rotation is thought to be the primary driver of dynamo action in these fully convective stars \citep{1995ApJ...453..464H, 2006A&A...454..889C}. By examining how rotation periods vary with mass and age in our sample, we can test and refine existing models of angular momentum evolution in low-mass stars and potentially uncover new insights into the underlying physical processes.

White-light flares, which exhibit broadband emission in ranges from the near ultra-violet through to optical wavelengths, sometimes extending into the far ultra-violet and near infra-red, have been studied extensively in recent years thanks to space-based dedicated photometric monitoring missions like the Kepler \citep{2010Sci...327..977B} and its extension NASA’s K2  \citep{2014PASP..126..398H}, and ongoing the Transiting Exoplanet Survey Satellite (TESS, \citealt{2015JATIS...1a4003R,2016AGUFM.P13C..01R}). We focus on M dwarfs in YMGs to investigate their flare activity, primarily using data from TESS. This survey provides high-cadence, long-duration light curves that are ideal for detecting and characterising stellar flares across a wide range of energies. By analysing the TESS data, we aim to establish the frequency distribution of flares for our sample of young M dwarfs, providing a comprehensive view of flare occurrence rates and energetics in these active stars.

To complement our analyses of the optical observations, we also examine far ultraviolet (FUV) data from the Cosmic Origins Spectrograph (COS) on the Hubble Space Telescope (HST) for a subset of our sample. FUV observations provide valuable insights into the high-energy radiation environment of M dwarfs, which is particularly relevant for assessing planetary atmospheric loss and photochemistry \citep{2005ApJ...622..629D, 2024ApJ...960...62E}. The FUV emission from M dwarfs, especially during flares, can have significant impacts on planetary atmospheres, potentially leading to atmospheric escape or the formation of prebiotic compounds \citep{2007AsBio...7..185L,2010AsBio..10..751S}.

Recent studies have challenged the accuracy of current flare models and UV emission predictions from optical data and have investigated flare frequency distributions (FFDs) in UV wavelengths. \citet{2019ApJ...871..167K} found that the Balmer continuum and line emission increased NUV flux 2-3 times above optical continuum extrapolations. \citet{2023ApJ...944....5B} identified short-duration NUV flares in GALEX data that lacked detectable optical counterparts in simultaneous Kepler observations. \citet{2024ApJ...971...24P} found that NUV flares occur with a considerably higher frequency compared to optical flares. \citet{2023ApJ...955...24R} noted that UV flare rates significantly exceed those observed by TESS and Kepler in visible/NIR wavelengths (white-light flares), with differences spanning several orders of magnitude. These findings could indicate differences in FFDs of the observed white light and UV flares.

By combining TESS optical data with HST COS FUV observations, we aim to provide a comprehensive view of the magnetic activity of young M dwarfs across different wavelength regimes. This multi-wavelength approach allows us to probe different layers of the stellar atmosphere and better understand the physical processes driving flare activity \citep{2021ApJ...915...37W}. Our study builds upon previous work on stellar activity in young stars \citep{1993ApJ...414L..49C, 2002A&A...390..219B} and extends it to a larger sample of M and late K dwarfs in well-defined YMGs. By focusing on these young systems, we hope to shed light on the young M dwarf magnetic activity and its implications for planetary formation,  evolution and habitability \citep{2002ApJ...581..626R, 2007AsBio...7..185L, 2010AsBio..10..751S, 2016ApJ...830...77V, 2019AsBio..19...64T}.
Understanding the evolution of magnetic activity in M dwarfs informs predictions of long-term planetary habitability \citep{Owen1980}, guiding future exoplanet surveys and prioritising targets for next-generation telescopes.

\section{Methods}
\label{sec:methods}

\subsection{The sample}

A primary goal of this paper is to understand the flare frequency distribution among young M dwarfs and relationship with stellar properties such as the period of rotation and the age. Clusters of stars that formed at the same time and are located close to each other serve as crucial test sites. These so-called YMG are kinematic associations of nearby stars that share a common origin and therefore age, which can be determined from isochrone dating. We include in our sample the following M and late K dwarf members of such groups in order from young to old: Greater Taurus Subgroup 8, Chamaeleon I, TW Hydrae, Lower Centaurus Crux, $\beta$ Pictoris, Argus, Columba, Tucana-Horologium, Carina, AB Doradus associations and more aged Hyades cluster (see Table \ref{tab:1}). In order to populate the sample with field stars, we consider the CARMENES M dwarf sample provided by \citet{2024A&A...684A...9S} and the multi-planet system's host stars sample from \citet{2024A&A...685A.143M}. From the latter sample, we made a separate selection of K and M-dwarf stars only (see Table \ref{tab:2}).

\begin{table}
\caption{\label{tab:1}Young moving groups in the sample}
   \centering

\small
   \begin{tabular}{c c c c c c c }
   \hline
   Association & Age (Myr) & Nstars & References  \TBstrut\\
    \hline
   Greater Taurus Subgroup 8 & 4.5 & 15& (1) \TBstrut\\
   \hline
    Chamaeleon I & 5 & 19& (2,7)\TBstrut\\
   \hline
    TW Hydrae & 8-20 & 7& (3,4) \TBstrut\\
   \hline
    Lower Centaurus Crux & 15 & 1 & (5)\TBstrut\\
   \hline
    $\beta$ Pictoris & 12-24 & 150& (3,4,9)\TBstrut\\
   \hline
    Argus & 30-50 & 10&(3,4,6)\TBstrut\\
   \hline
    Columba & 42 & 10&(3,5)\TBstrut\\
   \hline
    Tucana-Horologium & 45 & 122& (3,5,8)\TBstrut\\
   \hline
    Carina & 45 & 2& (3,4)\TBstrut\\
   \hline
   AB Doradus& 149 & 10& (3,5)\TBstrut\\
   \hline
   Hyades cluster& 800 & 1& (5)\TBstrut\\
   \hline
   \end{tabular}

   \tablefoot{Nstars is the number of stars in the sample belonging the YMG. Age references: (1) \citet{2021ApJ...917...23K}, (2) \citet{2007ApJS..173..104L}, (3) \citet{2017ApJ...846...93S}, (4) \citet{2013ApJ...762...88M}, (5) \citet{2018ApJ...862..138G}, (6) \citet{2019ApJ...870...27Z}, (7) \citet{2015A&A...575A...4F}, (8) \citet{2014AJ....147..146K}, (9) \citet{2019AJ....157..234S}.}
\end{table}

\begin{table}
\caption{\label{tab:2}Young and field stars in the sample}
   \centering

\small
   \begin{tabular}{c c c c c c c }
   \hline
   Subsample name & Age (Myr) & Nstars & References  \TBstrut\\
    \hline
   CARMENES & 25 & 5& (11) \TBstrut\\

    & 50 & 6& \TBstrut\\

 & 120 & 10& \TBstrut\\

& 800 & 34 & \TBstrut\\

    & >800 & 78& \TBstrut\\

    Multi-planet & $\le$800 & 5& (13)\TBstrut\\
    Multi-planet & >800 & 8&\TBstrut\\

   \hline
   \end{tabular}

   \tablefoot{The subsamples references : (11) \citet{2017ApJ...846...93S}, (12) \citet{2024A&A...685A.143M}.}
\end{table}

\subsection{Data collection and processing}
Flares are known to enhance chromospheric activity indicators such as equivalent width (EW) of Balmer H$\alpha$ \citep{2010AJ....140.1402H,2013ApJS..207...15K}. We retrieved H$\alpha$ emission data from literature and the Gaia astrometric space mission recent data release (Gaia DR3, \citealt{2023A&A...674A...1G,2023A&A...674A..28F}). For a small number of stars lacking literature values, we used Gaia's H$\alpha$ pseudo-equivalent width (pEW) as an estimate of H-alpha line strength, particularly for the youngest stars in our sample.

For YMG, it is a custom to report the age of the group and spectral types of the members, but other fundamental parameters such as mass, radius, and surface gravity of the star are rarely reported, requiring further characterisation of targets. For the sample stars, we obtained mean magnitudes in the G band and in the integrated RP band from the Gaia DR3 dataset, along with mean TESS T magnitudes for stars in our sample obtained from TESS Input Catalog (TIC, \citealt{ricker2014transiting}). Direct mass measurements for faint, isolated (non-binary) stars are often challenging \citep{2010A&ARv..18...67T,2019AJ....158..138S}. To obtain stellar masses and radii, and in some cases stellar effective temperature, we use absolute T magnitudes and the known ages fitting them into isochrones provided by MESA Isochrones and Stellar Tracks (MIST, \citealt{2016ApJS..222....8D,2019ApJS..243...10P}). The apparent G and RP magnitudes from Gaia DR3 were used for activity analyses.

This study sample stars were observed during the Kepler\footnote{http://www.nasa.gov/mission\_pages/kepler/overview} \citep{2010Sci...327..977B} and TESS\footnote{http://www.nasa.gov/tess-transiting-exoplanet-survey-satellite} \citep{2016AGUFM.P13C..01R} missions. We primarily utilised light curve data from the TESS mission, supplementing with data from the Kepler and K2 surveys in cases where TESS data was unavailable. We processed these data using the respective pipelines for each mission. Data was downloaded from the Mikulski Archive for Space Telescopes (MAST) in various cadences (for details, see Appendix \ref{sec:appendix1}). We used these light curves to determine rotation periods and analyse flare events.

\subsection{White-light light curve analysis}
\label{subsec:wlca}
We determined stellar rotation periods using long-cadence (LC) data from TESS, Kepler, and K2. The analysis was primarily conducted using the 'Lightkurve' module and the Lomb-Scargle method in the frequency domain.

Typically, we collected light curves from all quarters and analysed them one by one, in the frequency domain using the Lomb-Scargle method \citep{1976Ap&SS..39..447L,1982ApJ...263..835S,2018ApJS..236...16V}. We generally use the Pre-search Data Conditioned Simple Aperture Photometry (PDCSAP) fluxes in this study as they are the result of a systematic reprocessing of the entire Kepler and TESS database using a Bayesian approach to remove systematics from the short and long-term light curves. However, the reported periods were also reconfirmed in SAP fluxes as aforementioned reprocessing can remove stellar variability needed for determining periods \citep{2019MNRAS.489.5513C}. The Lomb-Scargle method works by fitting a sinusoidal curve at each of the frequencies in the periodogram and uses this fit to determine the value of power each frequency has in the periodogram.  In Appendix \ref{sec:appendix2}, we plotted Fig. \ref{fig:3}, and the panels from the left to the right show an example of a light curve, periodogram and folded curve of UCAC4 208-001676 (2MASS J01484771-4831156) in TESS Sector 2 using PDCSAP.

We focused our flare analysis on short cadence (SC) light curves from TESS and Kepler, using PDCSAP fluxes, as the SC data is most sensitive to flares \citep{2015ApJ...800...95L}. We implemented a newly developed Python routine "Young M Dwarfs Flares" ("YMDF") \footnote{\url{https://github.com/cepylka/ymdf}} based on the 'altaipony' module \citep{2021A&A...645A..42I,2022ascl.soft01009I} for finding and analysing flares in Kepler, K2, and TESS photometry. The flare finding routine 'flare finder' included in "YMDF" allows us to work with any time series comprised of flux data with corresponding errors, and we used it in subsequent analyses of far ultra-violet light curves.

To study flare frequency for the sample stars, we process the light curves and correct the detection efficiency following these steps for each star in our sample: (i) we de-trended the light curve, (ii) found flare events, (iii) performed tests of injection and recovery for flare events, and finally, (iv) corrected the number and energy of flare events based on the detection probability and energy recovery ratio obtained from step (iii), respectively.
For flare detection, we applied criteria based on \citet{2015ApJ...814...35C}, identifying significant statistical outliers 3$\sigma$ above the iterative median flux. To account for the potential effects of the 'flare finder' algorithm, we applied an injection-recovery routine to compare recovered events to injected ones. Only stars that exhibited more than three detected and successfully recovered flare events were included in the final sample. A detailed description of the routine is provided in Appendix \ref{subsec:absdetect}.

For every star in the sample, by running the injection-recovery routine, we collected all potential flare events and the following information: recovered amplitude measured relative to the quiescent stellar flux, recovered equivalent duration (ED) \citep{1972Ap&SS..19...75G} of the flare, the minimum uncertainty in ED derived from the uncertainty on the flare flux values, see Eq.\ref{eq:5} (\citealt[Eq.2]{2016ApJ...829...23D}), start and end of flare candidates in units of cadence, array index and actual time in days. The injection-recovery routine produced corrected values for amplitude, ED, ED standard deviation ($\sigma_{ED}$), duration and recovery probability with corresponding standard deviation.

The ED values provide relative energy for each flare event without having to flux calibrate the light curves. This is a universal parameter, which allows comparing flares in different stars and it is widely used in the literature (e.g. \citealt{2014ApJ...797..121H,2014ApJ...797..122D,2015ApJ...800...95L,2016ApJ...829...23D,2021A&A...645A..42I,2021A&A...650A.138S}). The ED is the area under the light curve with the quiescent flux subtracted, i.e. the time needed for the quiescent star to radiate the same amount of energy that was released during the flare event.
It can be formulated as the integrated flare flux divided by the median quiescent flux F$_\mathrm{0,\star}$ of the star, integrated over the flare duration \citep{2012PASP..124..545H}:

\begin{equation}\label{eq:5}
ED = \int dt \frac{F_{\mathrm{flare(t)}}}{F_\mathrm{0,\star}}
\end{equation}

The energy of the flare emitted in the Kepler/TESS bandpass (units of ergs) can be determined from the ED (units of seconds) by multiplying by the quiescent luminosity (units of erg s$^{-1}$). Flare energies were calculated as follows:

\begin{equation}\label{eq:6}
E_\mathrm{Kp/TESS,flare}= L_\mathrm{Kp,\star} \cdot ED
\end{equation}

The quiescent flux can be estimated if we assume blackbody radiation from the effective temperature T$_\mathrm{eff}$ and radius of the star R$_\star$:

\begin{equation}\label{eq:61}
 F_\mathrm{Kp/TESS,\star}=\int_{\lambda 1}^{\lambda 2} \mathcal{F}(\lambda) S_{\mathrm{res}}(\lambda)d\lambda,
\end{equation}
Here, $\mathcal{F}(\lambda)$ is the blackbody radiation at T$_\mathrm{eff}$ and $S_{res}(\lambda)$ is the Kepler/TESS response function \citep{2017ApJ...841..124V}. While the blackbody approximation omits molecular opacities inherent to more sophisticated models, its use preserves direct comparability with previous studies of stellar activity (e.g. \citealt{2013ApJS..209....5S,2019A&A...622A.133I,2021A&A...645A..42I,2023ApJ...945...61N}). The broad, red-sensitive TESS bandpass provides more robust photometric measurements for M dwarfs, whose complex molecular absorption could otherwise introduce significant errors in narrower or bluer bandpasses \citep{2015JATIS...1a4003R}.
We compute this photospheric specific flux, which is normalised to have the same count rate through the Kepler or TESS filters as the photosphere of each target. From that, we obtained  $L_\mathrm{Kp/TESS,\star}$, the the projected quiescent luminosity following \citet{2013ApJS..209....5S} and \citet{2019A&A...622A.133I}:

\begin{equation}\label{eq:7}
L_\mathrm{Kp/TESS,\star} = F_\mathrm{Kp/TESS,\star} \cdot \pi \cdot  R_\star^2,
\end{equation}
which allows us to calculate the energy of every flare exhibited by stars in our sample.

The flare frequency distribution (FFD) describes the rate of flares as a function of energy and follows the probability distribution:

\begin{equation}\label{eq:17}
N(E)dE = \beta E^{-(\alpha - 1)} dE,
\end{equation}

where $\alpha$ is the slope of the power law and $\beta$ is a normalisation constant.
In the cumulative form, the frequency of flares above a certain energy E$_\mathrm{cut-off}$ is defined as:

\begin{equation}\label{eq:18}
f (> E_\mathrm{cut-off}) = \frac{\beta}{\alpha - 1} E^{-\alpha+1}
\end{equation}

ED can be inserted in Eq.\ref{eq:17} and Eq.\ref{eq:18} instead of energy, and as the ED values are essential relative energy, and this formalism allows to study multiple stars in context of cumulative FFD calculations, assembling the subsamples based on certain properties of these stars (i.e. periods, mass, and age).

To analyse the flare frequency distribution (FFD), we used the Modified Maximum Likelihood Estimator (MMLE, \citealt{2009MNRAS.395..931M}) to find the slope $\alpha$, and least squares fitting to estimate the intercept $\beta$. These values were then used as initial guesses for assessing the power law distribution behaviour in the sample of flares.

\subsection{FUV light curve analyses in COS-HST}
\label{subsec:metoduv}

FUV observations reveal M dwarfs' high-energy radiation environment, and it is important in context of FUV emission, especially during flares, significantly impacting planetary atmospheres. Findings \citep{2019ApJ...871..167K,2023ApJ...944....5B,2023ApJ...955...24R} show that UV flare rates significantly exceed those observed in visible/NIR wavelengths (white-light flares) by several orders of magnitude, potentially indicating differences in FFDs between observed white light and UV flares.

We analysed archival Hubble Space Telescope (HST) data to identify low-mass stars belonging to our sample and previously observed using the Cosmic Origins Spectrograph (COS). This equipment, employed for far-ultraviolet (FUV) spectroscopy aboard HST, has been instrumental in studying FUV emission from stellar flares \citep{2012ApJ...750L..32F,2018ApJ...867...71L,2024MNRAS.533.1894J}. We focus our analysis on FUV data from COS's G130M diffraction grating (usually used for $\sim$117.0 – 143.0 nm observations), which includes strong transition region emission lines (C II, C III, Si III, Si IV, N V) and weaker lines. Ly$\alpha$ and OI are typically lost to geocoronal airglow. COS's photon-counting detector allows flexible binning of wavelength and time for creating light curves, integrated spectra, and subsampled spectra (although in practice, meeting S/N thresholds often requires coarser spectral or temporal resolutions).

We utilised existing observations to investigate the FUV characteristics of flare events in low-mass stellar targets with the search criteria focused on observations conducted in TIME-TAG mode, which records the arrival time and detector position of each photon with a typical precision of 1.25$\times$10$^{-4}$ seconds. This mode enables the construction of light curves with user-defined cadences.
The Cosmic Origins Spectrograph (COS) data reduction pipeline, CALCOS, processes raw HST/COS observations to produce calibrated spectra \citep{2023cosi.book...16H}. It performs wavelength and flux calibrations, corrects for instrumental effects, and generates one-dimensional spectral products.

\begin{figure*}[h!]
\sidecaption
\includegraphics[width=12cm]{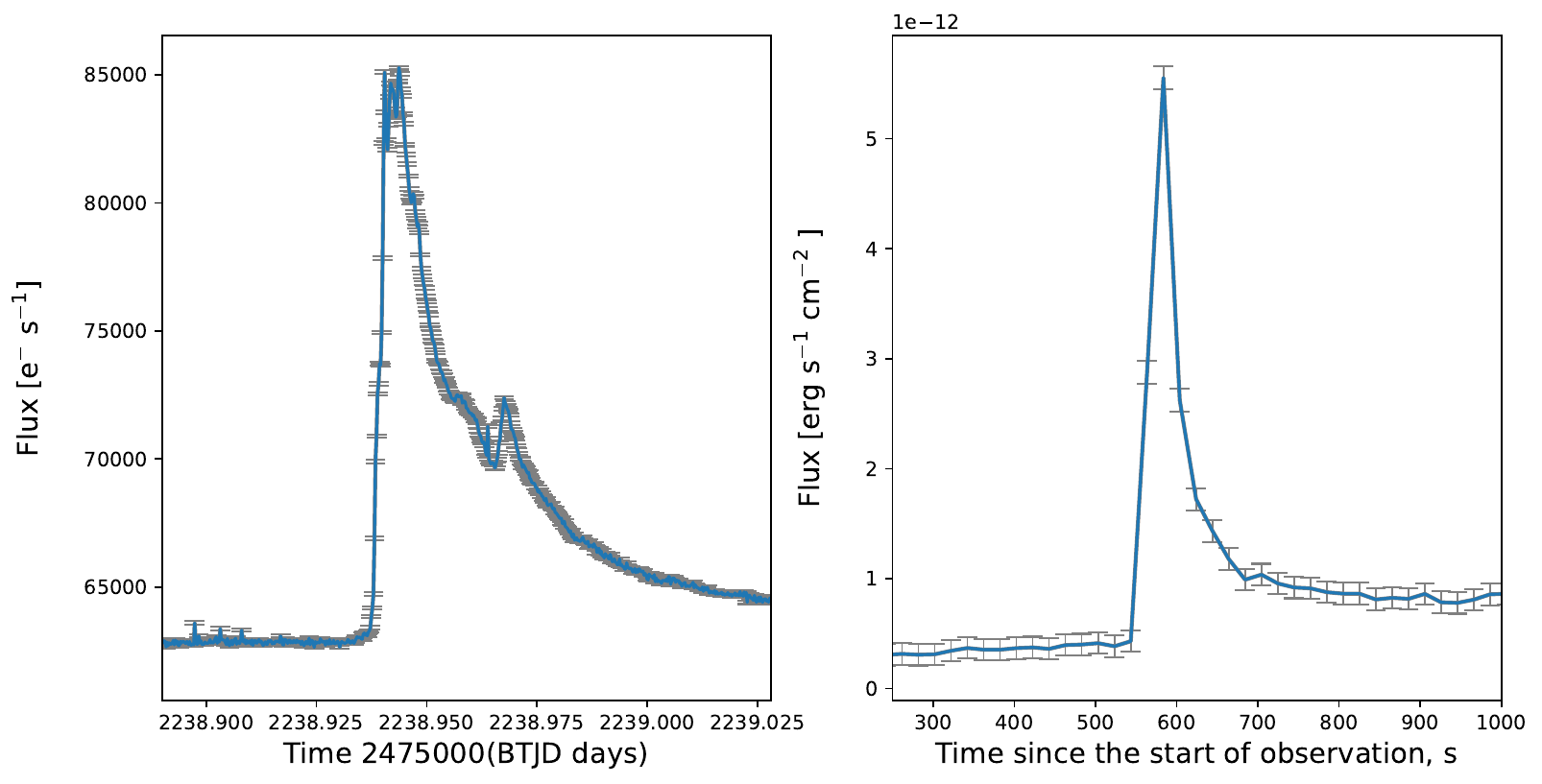}
\caption{Example of flares detected with the finding algorithm, described in Appendix \ref{subsec:absdetect}, shown for TESS (left) and COS (right) observations of Karmn J07446+035 (YZ CMi). The TESS flare, found in Sector 34, the fast (20 s) cadence observation has an ED=978.54 s and the total energy in TESS bandpass of $\approx$2.38$\times$10$^{33}$ erg. Applying a correction factor of 0.19 for the TESS CCD response and assuming a blackbody model of 9000 K \citep{2024MNRAS.527.8290P,2022ApJ...926..204H} to calculate the bolometric energy of the flare, we obtained  E$_{bol} \approx$1.3$\times$10$^{34}$ erg, consistent with \citet{2023ApJ...945...61N}.
The right panel shows a flare in COS-HST observed 2024-04-12 with starting time 08:36:35.26, with ED=1836.18 s and a total energy in G130M range of $\approx$2.70$\times$10$^{29}$ erg.}

\label{fig:2}
\end{figure*}

For flare finding, we again used the Python routine "YMDF", designed to analyse flux from various sources in the form of time series with associated uncertainties.
Fig. \ref{fig:2} shows the example of flares detected with the finding algorithm in TESS (left panel) and COS-HST (right panel) data.

In order to calculate the energy release in the obtained FUV range (106.0–136.0 nm excluding detector gap between 121.0–122.5 nm covering Ly$\alpha$) for a specific flare event, we use the quiescent surface flux of each flare star calculated from iterative median values of the flux between flare events (i.e. from light curve excluding outliers as was described above).

\begin{equation}\label{eq:86}
F_{\mathrm{int},\star}  = F_{obs} \cdot (D_\star / R_\star)^2,
\end{equation}

where $F_{obs} = \text{median}(F_{\mathrm{obs},\star}  \setminus \text{NaN})$ is the observed de-trended iterative median flux in the span of observation with excluded infinite values, $D_\star$ is the distance from the Sun and $R_\star$ is star radius.

\begin{equation}\label{eq:87}
L_\mathrm{FUV,\star} = F_{\mathrm{int},\star } \cdot \pi \cdot  R_\star^2
\end{equation}

Here, $L_\mathrm{FUV,\star}$ is the quiescent luminosity in the specific FUV range, $F_{\mathrm{int},\star }$ is the surface flux, and $\pi \cdot R_\star^2$ is the projected disk area of a star.

For our sample stars, we employed the stellar distances most recently reported in the literature and obtained via querying the SIMBAD Astronomical Database \citep{2000A&AS..143....9W}, as Gaia DR3 do not report distances for numerous targets in our study. The distance values found in the literature for nearby stars up to $\sim$2 kpc (in case of our sample <180 pc) exhibit precision comparable to Gaia benchmarks, as demonstrated by cross-validation studies (\citealt[Fig.~7]{2023A&A...674A..28F}). Additionally, we derived stellar radii using MIST isochrones, using the TESS T absolute magnitudes. The similar approach was suggested by \citealt{2018ApJ...860L..30H,2022ApJ...926..204H}. In \citet{2022AJ....164..110F}, the authors presented AU Microscopii panchromatic spectra with high resolution. We reconfirmed our approach using this panchromatic flux data for AU Microscopii and compared it with retrieved quiescent spectra values from 5 different observations of the star by COS-HST. We retrieved the mean value of 1.39$\times$10$^{28}$  erg within the wavelength range of G130M grating (105.815 - 137.439 nm with covered Ly$\alpha$), and the quiescent luminosity calculated from panchromatic spectra is similar (1.4$\times$10$^{28}$ erg s$^{-1}$ according to \citet{2022AJ....164..110F}). Therefore, we extended this method for the calculation of quiescent luminosity and released energy in this specific range to the other stars in the sample that have FUV observations.

\section{Results}
\label{sec:results}
\subsection{Stellar rotation period analyses}
\label{sec:resultsp}
\begin{figure*}\resizebox{\hsize}{!}{
   \centering
   \includegraphics{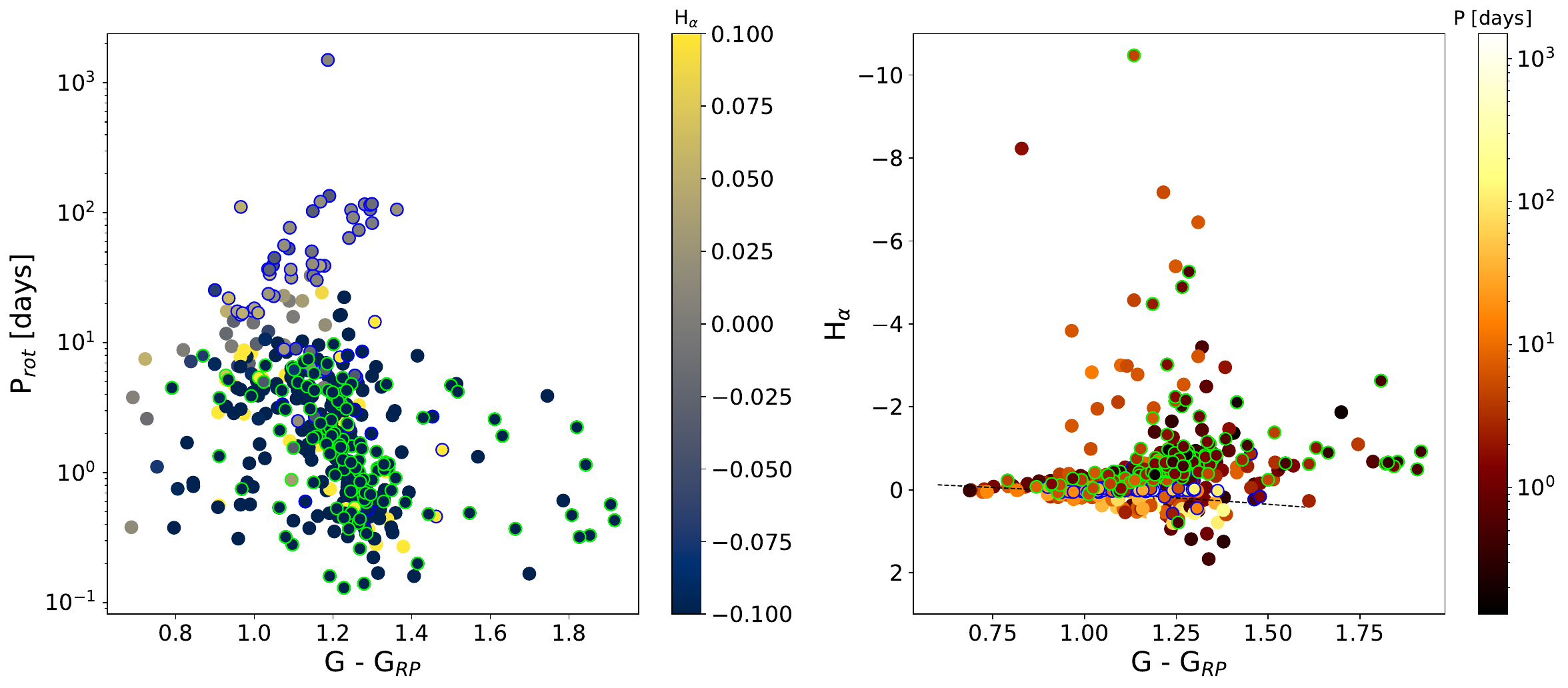}
    }
\caption{The relationship between H$\alpha$  and logP$_{rot}$ across (G - G$_{RP}$) colour from Gaia DR3.   Left: the rotation period distribution with (G - G$_{RP}$) colour-coded by normalised H$\alpha$ equivalent width. Right: H$\alpha$ equivalent width against (G - G$_{RP}$), colour-coded by logP$_{rot}$. Stars with newly determined periods P$_{rot}$ are plotted as circles with lime-green edges, field stars in the sample are represented as circles with blue edges. The M-dwarf activity boundary plotted as a dashed line follows \citet{2021AJ....161..277K}.}
\label{fig:12}
\end{figure*}

We reported newly determined rotation periods for 119 young stars, and the found periods span from 0.14 to 9.75 days. These stars are predominantly members of the $\beta$ Pictoris and Tucana-Horologium YMGs. While rotation periods can be challenging to measure, particularly with short-baseline data like TESS light curves, they are generally reliable for our sample. This is because our dataset is primary composed of young stars exhibiting short rotation periods, and relatively large modulation amplitudes, which are easier to identify and characterize, even within the constraints of TESS observational windows \citep{2024A&A...684A...9S}. Generally, rotation determination is performed using the PDCSAP data from long-duration space missions like Kepler and TESS, and it was our method of choice in this study.

We measured the period of rotation for the sample stars in order to distribute them in period bins in days: P$_{rot}$<0.6, 0.6$\leq$P$_{rot}$<1.85, 1.85$\leq$P$_{rot}$<4, 4$\leq$P$_{rot}$<7, 7$\leq$P$_{rot}$<30, P$_{rot}\geq$30. Our bin grid selection aimed to effectively distribute sample stars and isolate very fast rotators in the initial bin. We report found periods in the table with the main sample planet parameters available at the CDS. Many stars in the sample already have rotation periods reported in the literature, especially for the CARMENES sample \citep{2024A&A...684A...9S}, and if found values of rotation period were inside a 10\% interval from the literature values, we used our determined periods in subsequent analyses. Our period of rotation calculations sometimes provide different results from TESS and Kepler photometry, and also differ from those found in the literature. In this case, we exclude such stars from the sample. Some of the stars in the sample, whose periods were determined by the periodogram method, come out with a large discrepancy (>50\% difference in values derived for different sectors or from PDSSAP and SAP data. If such stars have the majority of their detected periods belonging to the same period bin, we included such stars in the sample, and we marked them (33 stars total) with a special uncertainty flag. We address the potential explanation of these discrepancies in Sect. \ref{sec:discussion}.

\begin{figure*}\resizebox{\hsize}{!}{
   \centering
   \includegraphics{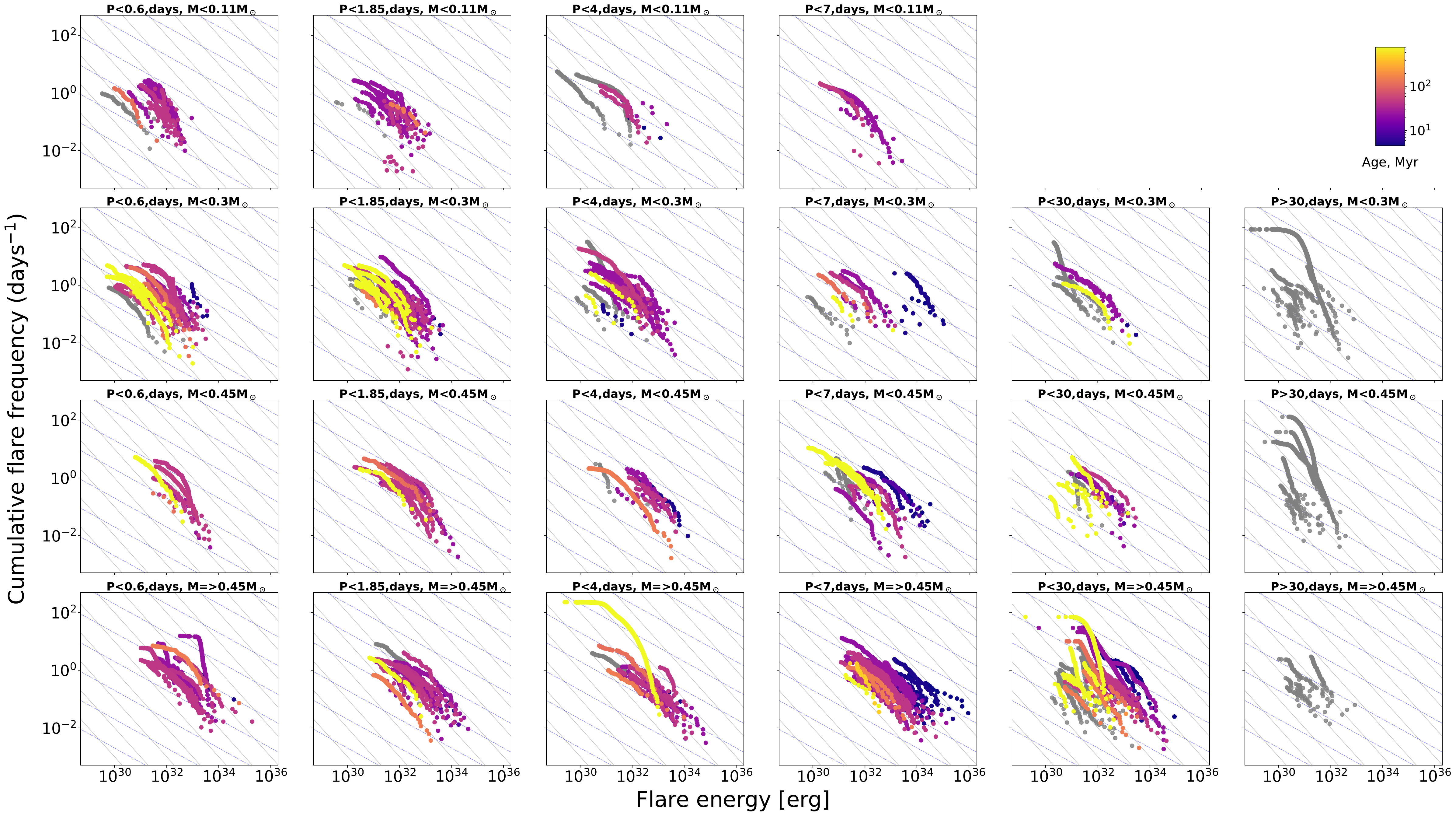}
    }
\caption{The distribution of the found flare populations in the sample stars in the cumulative form (FFD). The sample is binned in mass and period bins as described in Sect. \ref{sec:resultsp}. The ages are colour-coded from blue to red to yellow for 4.5-800 Myr; field stars with age > 800 Myr are plotted as grey circles. The grey and blue dashed guides corresponding to power law coefficient $\alpha$=2.0 and $\alpha$=1.5, respectively, for a range of flare energies $\in$[10$^{30}$, 10$^{36}$] erg in TESS bandpass.
}
\label{fig:16}
\end{figure*}

In Fig. \ref{fig:12}, we plotted the relationship between H$\alpha$  and logP$_{rot}$ across (G - G$_{RP}$) colour. In these plots, we distinguish between young stars and field stars (> 800 Myr). The equivalent width of the H$\alpha$ emission line for the young stars increases with redder values of (G - G$_{RP}$). The observed variation is primarily attributed to fluctuations in the photosphere's contribution rather than an increase in magnetic activity (\citealt{1986ApJS...61..531S}). Many YMG stars in the colour range corresponding to M dwarfs $\sim$(G - G$_{RP}) \gtrsim$0.9 exhibit high H$\alpha$ activity (they fall above the activity boundary line \citep{2021AJ....161..277K}, as expected for their young age \citep{2021ApJ...916...77P,2023ApJ...945..114P}. The authors note that this could signal that at the age of Tucana-Horologium, YMG H$\alpha$ values are independent of the rotation period.

We observe a subtle correlation where stars with shorter rotation periods and redder (G - G$_{RP}$) colours tend to exhibit slightly larger H$\alpha$ equivalent widths, though, curiously, this does not always apply to the fastest rotators (P$_{rot}$<0.6). This trend is consistent with the relationship between stellar rotation, colour, and chromospheric activity for young M dwarfs \citep{1972ApJ...171..565S,2008ApJ...687.1264M}.  We also comment that the newly determined periods (plotted as circles with lime-green edges) seem to align with the properties of the other young stars in the sample, and exhibit a similar relation between age, period and chromospheric activity indicators. The field stars plotted as circles with blue edges exhibit less activity in H$\alpha$.

We also examine the rotation period distribution in the context of stellar mass, which we determined based on fitting MIST isochrones to the TESS T absolute magnitudes (see Sect. \ref{sec:methods}). We binned our sample as follows: M$_\star$<0.11M$_\odot$, 0.11M$_\odot$$\leq$M$_\star$<0.3M$_\odot$, 0.3M$_\odot$$\leq$M$_\star$<0.45M$_\odot$, M$_\star$$\geq$0.45M$_\odot$. We aimed to populate bins effectively, noting that the first two bins likely contain only fully convective stars. The value of T magnitude for the faintest stars above the values in MIST isochrones for various young ages informed our choice for the right edge of the first bin, and we note that sometimes the magnitudes reported in TESS are too faint to correspond to even the smallest mass value for a certain age. This approach enables distinct analysis of late and mid-late M dwarfs compared to early M and K dwarfs while maintaining modest-sized subsamples in each bin.

Another useful parameter for describing stellar magnetic activity-rotation relations is the Rossby number, defined as Ro=P$_{rot} / \tau$ where $\tau$ is the convective turnover timescale for a given stellar mass. We calculated it for the stars in the sample from the empirical log $\tau$ following \citet{2018MNRAS.479.2351W}. We used these numbers to discuss the implications of measured activity indicators in connection with laws governing frequency flare distributions in the Sect.\ref{sec:discussion}.

\subsection{Flare distribution analyses}
\label{sec:resultsffp}

We analysed flaring activity in the sample stars, following the procedure we described in Appendix \ref{subsec:absdetect}. We validated 86714 flare events with recovery probability >25\%. All events meet our criteria (see Eq. \ref{eq.1}, \ref{eq.2}, \ref{eq.3}, \ref{eq.4}) and were verified through the injection-recovery procedure, as described in Appendix \ref{subsec:absdetect}. The maximum, minimum and average flare energies detected in the sample are \mbox{E=9.23$\times$10$^{35}$} erg, \mbox{E=2.87$\times$10$^{28}$} erg, and \mbox{E=2.33$\times$10$^{32}$} erg, respectively which are the TESS flare energies calculated from the observed luminosity of a quiescent star (Kepler flare energies were converted to TESS bandpass using Equation \ref{eq.10}). The individual stars in our sample exhibited from 3 to 15969 flare events in the span of observation.

Using the YMDF routines, we look at the flaring rate, the observed energy released in flares, and the power law fit exponent $\alpha$ and intercept $\beta$ to the flare frequency distributions (FFDs). These interconnected flare metrics are often used as activity indicators. We proceeded to analyse flare activity by calculating FFDs for each star separately. We plot our resulting FFDs for calculated energies in TESS bandpass released during the flare in Fig. \ref{fig:16} in mass and period bins described above and provide grey dashed and blue dashed guides corresponding to power law coefficient $\alpha$=2.0 and $\alpha$=1.5, respectively, for a range of $\beta$ coefficients.
These guide lines facilitate comparison with values for single power law fits found by previous authors. For example, \citet{2020ApJ...905..107M} computed a weighted average of the $\alpha$=1.98 $\pm$0.02 for their sample of mid-to-late M dwarfs. \citet{2021A&A...645A..42I}, investigated K2 light curves of K and M stars and posit that $\alpha$ may be the same for all stars; a value of approximately 2. In the recent study, \citet{2024AJ....168...60F} analysed young stars (ages<300 Myr) and found $\alpha \sim$ 1.6 to 1.2 in TESS observations.

The youngest stars in our sample, aged 4.5–5 Myr, exhibit the highest flare energies across all mass ranges. This trend is particularly evident for stars with rotation periods between 4 and 7 days, meaning they are not the fastest rotators, consistent with expectations from being in the disc-locked phase prior to the dissipation of their protoplanetary discs (e.g. \citealt{2009IAUS..258..363I}). In contrast, the M dwarfs rotating faster have likely decoupled from their protoplanetary discs due to gas dissipation, enabling faster rotation. The presence or absence of these discs also has implications for ongoing planet formation processes around these young stars.

Stars aged 24-45 Myr populate all bins except the longest period, displaying a shallower slope ($\alpha \approx$ 1.5) for rotation periods of 1.85-7 days, resulting in a smooth distribution of flare energies.
Very low mass stars (M$_\star$<0.11 M$_\odot$) do not rotate slower than 7 days. These stars and the very fast rotators (P$_\mathrm{rot}$ < 0.6 days) comprise two populations that exhibit steeper power-law slopes, indicating frequent low-energy flares and rare super-flares, which is evident for young populations and some of the field stars. The picture changes when either mass or rotation period rises, and then the population leans towards a power law coefficient $\alpha$=1.5 up until when stars rotate slower than 7 days.

The behaviour of FFDs in the middle part of our grid could reveal the underlying connection between the mass of a star and corresponding period and activity indicators, such as the frequency of small, mid-size and extreme flares. The slope $\alpha$=1.5 implies that a star would not produce as many small-scale flares as would a star that obeys $\alpha$=2.0. However, there will be a rise in the number of mid-size flares and extreme flares in comparison. As to the largest events, it seems that there is no evidence of enhanced super-flare activity in a specific range of M dwarfs with 0.6$\leq$P$_{rot}$<7 days and 0.109M$_\odot$$\leq$M$_\star$, moreover, all stars at a certain very high energy cut-off will obey the larger slopes. Our results in TESS observations show that both populations in this range: old and young, show elevated activity in the middle-size flaring compared to $\alpha$=2.0. Fig.~\ref{fig:16} reveals diverse flaring energies among stars with comparable periods and masses. Notably, the FFD slopes of field stars generally align with those of young stars within the same bin.

From our results, it is apparent that stellar age is not the sole parameter governing flare behaviour. Although \citet{2024AJ....168...60F} found that flare rates are higher when stars are young (less than 50 Myr) and decrease with age, this does not apply to all M dwarfs. For example, the FFD for TRAPPIST-1, the old, low-mass (M$_\star$=0.08 M$_\odot$) and very active M7.5V dwarf with the value of $\alpha$=1.42 computed using MMLE method show similarities with the FFD for AU Microscopii, a young (belongs to $\beta$ Pictoris Moving group), more massive (M$_\star$=0.5 M$_\odot$) M0.5-1 pre-main sequence dwarf with power law coefficient of $\alpha$=1.46. These stars appear to occupy opposite extremes in stellar properties, yet exhibit similarly scaled FFDs when normalised to their respective energy regimes. Their derived rotation periods (3.3 days for TRAPPIST-1 and 4.83 days for AU Microscopii in this study) show close alignment with literature values of 3.304 days \citep{2019A&A...621A.126D} and 4.8 days \citep{2024A&A...691A.304Y}, respectively, and also do not differ greatly from each other.

\begin{figure*}\resizebox{\hsize}{!}{

   \centering
   \includegraphics{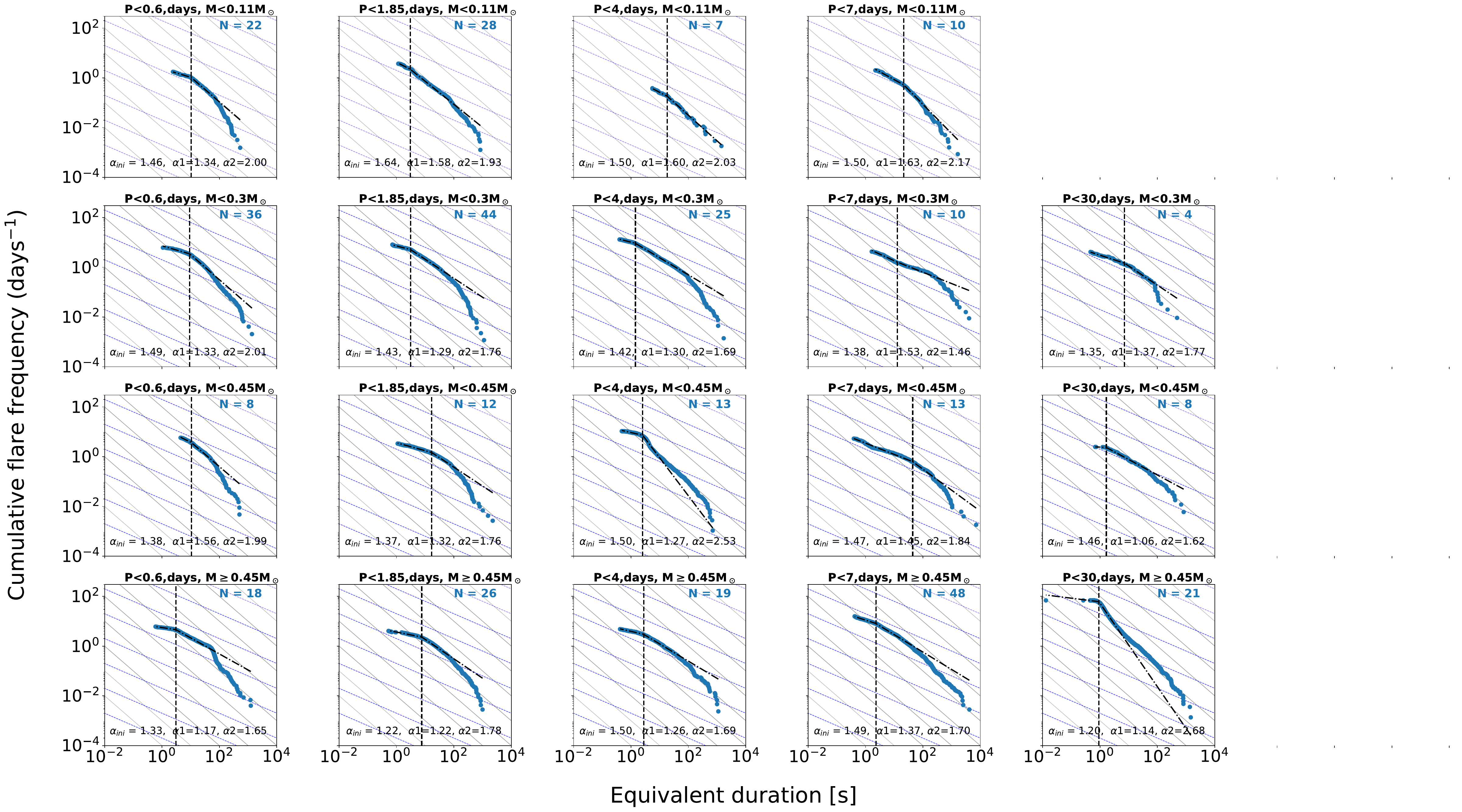}
    }
\caption{The cumulative distribution plotted for the mass-period bins for young (age < 800 Myr) stars in the sample. Blue dots represent the bin subsample FFD, which combines multiple stars by averaging each portion of the FFD by the number of stars that contribute to it.
We first calculated initial guess for slope $\alpha_{1~ini}$ and intercept $\beta_{ini}$ using MMLE method \citep{2009MNRAS.395..931M}. The initial guess for $\alpha_{2~ini}$ is fixed at 2.0. Then the broken power law was fitted to the distribution, resulting slopes $\alpha_1$, $\alpha_2$ (black dash-dotted lines) and break points (black dashed lines), if the fit was successful. The grey and blue dashed guides corresponding to power law coefficient $\alpha$=2.0 and $\alpha$=1.5, respectively, plotted for a range of ED $\in$[10$^{-2}$, 10$^{4}$] days.}
\label{fig:15}
\end{figure*}

We proceed further with analysing the binned sub-samples using the obtained and corrected equivalent durations of all flare events. Being a luminosity-normalised parameter, ED allows the comparison of flares in different stars, and the simultaneous analysis of flares taking into account the number of stars in sub-samples with different detection thresholds. To do this, we constructed multi-star FFDs, that is, summing the FFDs of all stars belonging to each mass-period bin and normalising the distribution at each ED value by the number of stars that contributed to it. In Fig. \ref{fig:15}, we plot the multi-star FFDs for young stars in our sample.  Blue dots represent corrected ED and the corrected individual frequencies of each flare by recovery probability. For these multi-star FFDs, we obtained power law coefficients $\alpha$ and intercepts $\beta$ using the MMLE method as before.

The physical meaning of the FFD power law slope has been discussed previously by several authors. For example, \citet{2017ApJ...841..124V} found $\alpha$=1.59 for TRAPPIST-1 and concluded that flare energies of this star are mostly nonthermal and are similar to the energies of other very active M dwarfs found by \citet{2014ApJ...797..121H}. \citet{2019A&A...622A.133I} noted that flare production is a process with self-similarity in the released energy patterns as it follows a power law; therefore, a deviation from this power law would reflect the underlying physical phenomena. \citet{2016ApJ...832...27A} argued that magnetic and thermal energies dominate around $\alpha \sim$ 2.0 as opposed to nonthermal energy around $\alpha \sim$ 1.4.  If both thermal and nonthermal processes are in play, it could be interesting to explore the idea that the flare activity will be better represented by the broken power law relation.
Based on spotted abrupt changes in power-law slopes in our sample binned for stars' masses and periods of rotation, we implemented a more sophisticated approach and considered the calculated values of $\alpha$ and the intercept $\beta$ from MMLE  as a first guess for a broken power law routine.

To model the broken power law relationship in our sample FFDs, we employed the BrokenPowerLaw1D class from the Astropy modelling package \citep{2022ApJ...935..167A}. This one-dimensional broken power law model is defined by four key parameters: amplitude $A$, x$_{break}$, $\alpha_1$, and $\alpha_2$. The amplitude parameter represents the model amplitude at the breakpoint, while x$_{break}$ denotes the location of the breakpoint. The model function is mathematically expressed as:

\begin{equation}\label{eq:17}
f(x) = \left \{
         \begin{array}{ll}
           A (x / x_{break}) ^ {-\alpha_1+1} & : x < x_{break} \\
           A (x / x_{break}) ^ {-\alpha_2+1} & :  x > x_{break} \\
         \end{array}
       \right.
\end{equation}

We initialize the BrokenPowerLaw1D model using MMLE-derived $\alpha$ for $\alpha_1$ and 2.0 for $\alpha_2$.  We propose that the break likely occurs at ED$_\mathrm{break}$=10 s. We then fit this model to our data using Astropy's LevMarLSQFitter or, if that fails, SimplexLSQFitter. If both fitters fail, we report only the initial power law coefficient estimates. The obtained $\alpha_1$, $\alpha_2$ and the breakpoints are stated in Fig. \ref{fig:15} and can be found in Table \ref{tab:3}.

The young stellar population in our sample generally exhibits initial FFD slopes, $\alpha_1$ $\lesssim$ 1.5, except for the lowest-mass stars, which may show slightly larger values. Regarding the second FFD slope, $\alpha_2$, we observe the following trends. The fastest rotators exhibit $\alpha_2$ values closer to 2.0, up to the highest mass bin. In contrast, the remainder of the sample demonstrates substantially shallower slopes. Additionally, the lowest mass bin also shows $\alpha_2$ values approaching 2.0.

We repeat our analysis on the whole sample (i.e., including both young and field stars) to assess how it could affect the slopes of FFDs, as plotted in Fig.~\ref{fig:25} in Appendix \ref{sec:appendix2}.
Notwithstanding the evident paucity of slowly rotating young stars, we observe minimal divergence between the two datasets, with the exception of the scarcely populated P$_\mathrm{rot} \in$[1.85,4] days and M$_\star$<0.11 M$_\odot$ bin, where TRAPPIST-1 flare events cause significant FFD deviation of $\alpha_2$ < 1.5. Field stars generally follow young star behaviour but predominantly rotate slower than 30 days, showing steep FFD slopes with low-energy flares in this regime. While our sample lacks young stars with P$_\mathrm{rot}$>30 days, the slow-rotating field stars maintain slopes $\alpha_2 \geq $2.0, possibly indicating the eventual fate of all M dwarfs at advanced ages, except for the least massive ones. We observe that the highest flare frequency shown in the multi-star FFD exhibits steeper FFD slopes, which correspond to fewer high-energy flares. The young population of the stars in the sample exhibits comparable distribution forms and analogous broken power law coefficients as the whole sample, despite adding the 85 field stars and 35 stars with an age of 800 Myr. Therefore, age does not appear to be an independent factor in determining the flare activity in M dwarfs. The rotational period appears to be a more fundamental indicator of the nuanced FFD behaviour of M dwarfs. Older stars with relatively rapid rotation (e.g., TRAPPIST-1) exhibit the same piecewise and shallower power law as younger stars in the same mass-period bin. Additionally, both fast rotators (<0.6 days) and slow rotators ($\geq$30 days) tend to have steeper $\alpha_2$ slopes compared to stars with intermediate rotation periods, regardless of age and up to the highest masses.

The high energy tails of the FFD deviate from $\alpha_2~$=2 power law towards steeper values for active stars in our sample, indicating that super-flares are even rarer than previously thought by many authors whose work we mentioned above. These deviations are prominent in all distributions, both for individual stars and the entire sample as seen in Fig. \ref{fig:16} and Fig.~\ref{fig:15}. We further discussed our findings in Sect. \ref{sec:discussion}.

\begin{table}
\scriptsize
\caption{\label{tab:3}Broken power law coefficients for the mass-period binned sample of young stars}
\centering
\resizebox{0.45\textwidth}{!}{\begin{tabularx}{9cm}{l | c c c c} \hline
\multicolumn{1}{l}{}&\multicolumn{4}{c}{M$_\star$<0.11M$_\odot$}\TBstrut\\
\hline
Period range (days) & $\alpha_{1~ini}$ & $\alpha_1$ & $\alpha_2$ & x$_{break}$ in ED (s)   \TBstrut\\
 \hline
P$_{rot}$<0.6      & 1.459 & 1.345$\pm$0.004 & 1.998$\pm$0.004 & 10.360$\pm$0.063 \TBstrut\\
0.6<P$_{rot}$<1.85 & 1.644 & 1.580$\pm$0.007 & 1.928$\pm$0.005 & 3.012$\pm$0.049 \TBstrut\\
1.85<P$_{rot}$<4.0 &1.500 & 1.601$\pm$0.012 & 2.032$\pm$0.021 & 18.174$\pm$0.674 \TBstrut\\
4.0<P$_{rot}$<7.0  &1.500 & 1.633$\pm$0.002 & 2.171$\pm$0.018 & 21.365$\pm$0.381 \TBstrut\\
\hline
\multicolumn{1}{l}{}&\multicolumn{4}{c}{\ 0.11$\leq$M$_\star$<0.3M$_\odot$} \TBstrut\\
 \hline
P$_{rot}$<0.6      & 1.491 & 1.331$\pm$0.003 & 2.007$\pm$0.004 & 9.244$\pm$0.048 \TBstrut\\
0.6<P$_{rot}$<1.85 & 1.435 & 1.294$\pm$0.003 & 1.765$\pm$0.002 & 3.109$\pm$0.019 \TBstrut\\
1.85<P$_{rot}$<4.0 & 1.420 & 1.300$\pm$0.001 & 1.687$\pm$0.001 & 1.427$\pm$0.004 \TBstrut\\
4.0<P$_{rot}$<7.0  & 1.382 & 1.527$\pm$0.005 & 1.455$\pm$0.010 & 12.825$\pm$2.820 \TBstrut\\
7.0<P$_{rot}$<30.0 & 1.354 & 1.374$\pm$0.005 & 1.771$\pm$0.032 & 7.107$\pm$0.478 \TBstrut\\
 \hline
\multicolumn{1}{l}{}&\multicolumn{4}{c}{0.3$\leq$M$_\star$<0.45 M$_\odot$}\TBstrut\\
 \hline
P$_{rot}$<0.6      & 1.376 & 1.560$\pm$0.007 & 1.991$\pm$0.005 & 10.601$\pm$0.112 \TBstrut\\
0.6<P$_{rot}$<1.85 & 1.374 & 1.316$\pm$0.002 & 1.761$\pm$0.006 & 16.616$\pm$0.186 \TBstrut\\
1.85<P$_{rot}$<4.0 & 1.500 & 1.268$\pm$0.003 & 2.526$\pm$0.009 & 2.536$\pm$0.010 \TBstrut\\
4.0<P$_{rot}$<7.0  & 1.471 & 1.454$\pm$0.003 & 1.844$\pm$0.060 & 44.210$\pm$4.267 \TBstrut\\
7.0<P$_{rot}$<30.0 & 1.459 & 1.057$\pm$0.018 & 1.623$\pm$0.004 & 1.657$\pm$0.024 \TBstrut\\
 \hline
\multicolumn{1}{l}{}&\multicolumn{4}{c}{M$_\star\geq$0.45M$_\odot$} \TBstrut\\
 \hline
P$_{rot}$<0.6      & 1.326 & 1.171$\pm$0.004 & 1.645$\pm$0.002 & 3.067$\pm$0.022 \TBstrut\\
0.6<P$_{rot}$<1.85 & 1.224 & 1.225$\pm$0.002 & 1.782$\pm$0.004 & 7.572$\pm$0.054 \TBstrut\\
1.85<P$_{rot}$<4.0 & 1.500 & 1.264$\pm$0.002 & 1.693$\pm$0.002 & 2.762$\pm$0.018 \TBstrut\\
4.0<P$_{rot}$<7.0  & 1.492 & 1.368$\pm$0.001 & 1.702$\pm$0.001 & 2.309$\pm$0.010 \TBstrut\\
7.0<P$_{rot}$<30.0 & 1.200 & 1.145$\pm$0.002 & 2.679$\pm$0.002 & 0.922$\pm$0.001 \TBstrut\\
\end{tabularx}}
\end{table}

\subsection{Flares in FUV}
\label{sec:resultsfuv}

We analysed flaring activity in the sample stars using the archival observations in FUV for active 11 young stars and 6 field stars, following the procedure we described in Sect. \ref{subsec:metoduv}. We found and validated 52 flare events in total (34 and 18 in the young and field stars, respectively). All events met the same criteria as for the TESS/Kepler events (see Eq. \ref{eq.1}, \ref{eq.2}, \ref{eq.3}, \ref{eq.4}) and were verified through the injection-recovery procedure. The average flare energy detected in the sample is $\sim$ 3.5$\times$ 10$^{31}$ erg, which is the flare energy calculated as E$_\mathrm{FUV}$=$L_\mathrm{FUV,\star} \cdot$ ED and we use Eq.\ref{eq:86}, \ref{eq:87} for the calculation of $L_\mathrm{FUV,\star}$. These results, along with the summary of the TESS observations for these stars, are presented in Table \ref{tab:4}.

\begin{table*}
\scriptsize
\caption{\label{tab:4}UV and white-light activity for several stars in the sample}
\centering
\resizebox{0.99\textwidth}{!}{\begin{tabularx}{20cm}{l c c c c| c c c c |c c c l} \hline
\multicolumn{5}{l}{}&\multicolumn{4}{c}{UV flares}&\multicolumn{4}{c}{TESS flares}\TBstrut\\
\hline
ID& SpT& P$_\mathrm{rot}$& T$_\mathrm{eff}$& Age (Myr)& n & ED (s)& E$_\mathrm{G130M}$& T$_{obs}$ (s)& n & ED (s)& E$_\mathrm{TESS}$ & Rate E$_{log31}$ \TBstrut\\
\hline
2MASS J11173700-7704381 &M0.5 & 12.8 & 3778 & 5.0& 2& 1.66-19.41 &0.19-2.23$\times$10$^{30}$ &4802.368 &92 &1.7-179.6 &0.04-1.32$\times$10$^{35}$ &141.7 \TBstrut\\
AU Microscopii$^1$ &M1 &4.8& 3835 & 24.0&14& 0.44-751.6& 0.002-2.50$\times$10$^{30}$& 42255.936& 202 &
0.42-134.6 &  0.013-4.12$\times$10$^{33}$& 14.3\TBstrut\\
2MASS J18141047-3247344$^2$&K5.0&1.71&3512&24.0&2&4.7-28.9& 0.10-4.77$\times$10$^{31}$&4503.744&36&2.83-569.94& 0.03-6.25$\times$10$^{32}$&0.62\TBstrut\\
2MASS J02365171-5203036$^3$& M2 &0.74&3626 &42.0 &4&24.35-8448.4&0.008-1.82$\times$10$^{31}$&9956.704&137&
2.39-702.27&0.02-6.35$\times$10$^{33}$&3.26\TBstrut\\
2MASS J01521830-5950168$^3$&M2&6.5&3200&45.0& 2&25.5-576.9& 2.53-5.00$\times$10$^{28}$ &3119.776 &161& 1.73-179.6&0.048-4.85$\times$10$^{32}$ &2.49
 \TBstrut\\
HIP 107345 &M1 & 4.5 & 3837 & 45.0& 2& 22.1-80.58 &0.20-1.24$\times$10$^{30}$ &5201.376 &110 &1.2-549.7 &0.004-1.67$\times$10$^{33}$  &1.9 \TBstrut\\
HIP 1993 &M0 &4.3 &4053& 45.0 & 1 & 253.9& 5.08$\times$10$^{28}$& 5369.376& 198&1.6-1238.0 &0.0005-4.18$\times$10$^{33}$ &1.58\TBstrut\\
2MASS J03315564-4359135$^3$& M0 &2.9& 4491& 45.0&1& 72.8& 1.85$\times$10$^{29}$&7413.408&51&2.05-49.37&0.13-3.17$\times$10$^{32}$& 1.36  \TBstrut\\
2MASS J02001277-0840516$^3$& M2.8 &2.28& 3345&45.0 &1&61.6 &1.22$\times$10$^{30}$&1760.85& 17&7.26-55.26&0.12-8.52$\times$10$^{31}$& 2.21 \TBstrut\\
2MASS J22025453-6440441$^3$&M2&0.43&3044&45.0& 1&267.8&2.85$\times$10$^{29}$&1332.0&109&1.91-535.86&0.002-5.90$\times$10$^{32}$&1.68\TBstrut\\
Karmn J07446+035$^{2,5}$& M4.5&2.78&3099&50.0 &4&115.09-288.12&0.02-2.66$\times$10$^{32}$&9586.688&277&0.43-978.54&0.001-2.37$\times$10$^{33}$&8.22\TBstrut\\
2MASS J17283991-4653424$^4$&M2.0&32.9& 3462& field &4 & 122.07-1259.76&0.1-1.11$\times$10$^{29}$& 12946.624& 65&0.18-49.58&0.025-7.87$\times$10$^{32}$ &2.62\TBstrut\\
Karmn J13102+477&M5.0&29.06&3221&field&4&28.16-626.77 &0.17-4.83$\times$10$^{28}$&13344.540&59&4.38-474.52&0.03-3.00$\times$10$^{32}$&0.46\TBstrut\\
Karmn J22096-046 &M3.5 &39.2&3229&field&6& 298.5-19465.7& 1.22-7.17$\times$10$^{28}$& 12640.800& 11&
0.27-3.33&0.19-2.32$\times$10$^{31}$&1.68\TBstrut\\
Karmn J16581+257 &M1.0 &23.8&3734&field&1& 215.1& 3.66$\times$10$^{28}$& 12638.24& 21&
0.27-3.13&0.19-2.32$\times$10$^{31}$&1.68\TBstrut\\
LTT 1445 A &M3.0 &1.41&3283&field&2& 1232.48-7325.77& 0.02-7.17$\times$10$^{30}$& 17654.752& 28&
2.2- 31.9&0.32-3.745$\times$10$^{31}$&5.78\TBstrut\\
TRAPPIST-1 &M8.0 &3.3&2566&field&1& 227.88& 1.66$\times$10$^{30}$& 12403.90& 232&16.3-1957&0.01-1.46$\times$10$^{32}$&0.29\TBstrut\\
\end{tabularx}}
   \tablefoot{Rate E$_{log31}$ is the rate in days$^{-1}$ for occurrence of flare with the energy E$_\mathrm{Kp/TESS,\star}\geq$ 10$^{31}$ erg. Several flares observed in FUV in our sample (n stands for the number of flares observed in FUV or TESS bandpass), discussed in literature: (1) \citet{2022AJ....164..110F}; (2) Mamonova et al.,in prep.; (3) \citet{2018ApJ...867...70L}; (4) \citet{2019ApJ...871L..26F}; (5) This study.}
\end{table*}

\begin{figure*}\resizebox{\hsize}{!}{
   \centering
   \includegraphics{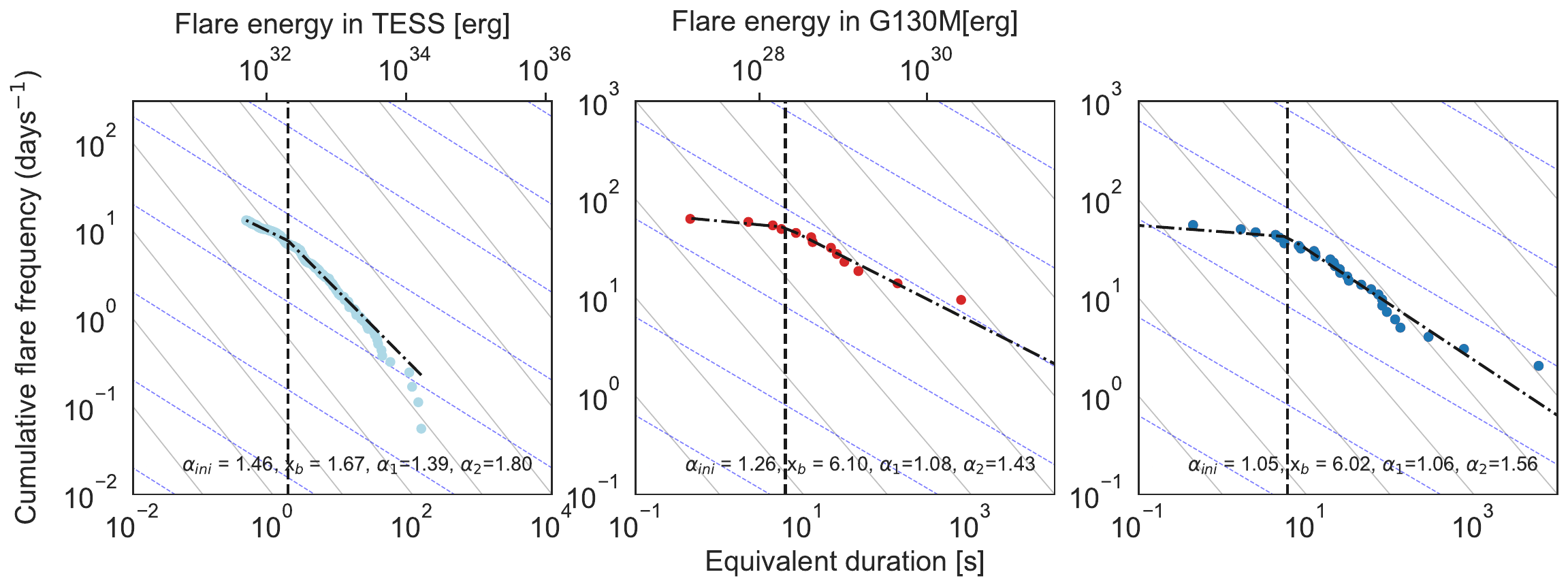}
    }
\caption{ Left and centre panel: Cumulative FFDs (scatter) in ED, and respective broken power law fits (dashed black lines). The y-axis shows the frequency distribution plotted against ED on the x-axis. The secondary x-axis on top represents flare energy. The left panel shows FFD in TESS flares for AU Microscopii. The centre panel represent FFD for FUV flares observed in the same star. Right panel: Cumulative FFDs (scatter) in ED, and respective broken power law fits (dashed black lines) for the sample of FUV flares in young M dwarfs. The y-axis shows the frequency distribution plotted against ED on the x-axis. For all panels: the grey and blue dashed guides corresponding to power law coefficient $\alpha$=2.0 and $\alpha$=1.5, respectively, for a range of ED/energies.}
\label{fig:23}
\end{figure*}

Only one star in our small sample of FUV flares, AU Microscopii, has them observed in sufficient amount for frequency analyses. This star exhibited 14 flares with a mean energy of 4.1$\times$ 10$^{29}$ erg in the specific FUV range during 11.7 hours of observation in total. In Fig. \ref{fig:23}, we plotted FFD in FUV flares (the left panel) and in TESS flares (the centre panel) with corresponding broken power law fits and coefficients. We plotted the fitted broken power law slopes, and for the FUV flares, the relations are less steep than for TESS flares; the flares themselves are less energetic and more frequent than those observed in TESS photometry. The high-end tail prominent in TESS FFDs is barely seen in the FUV data, but it can be explained by the lack of super-flares in the observations and the modest size of the sample.

As we obtained the corrected equivalent duration of all flare events in FUV, which allows us to compare flares in different stars, we proceed to analyse all FUV sample flares simultaneously. To construct the FFD, we employed the same methodology that incorporates the corrected ED values and the corrected individual frequencies. We applied recovery probability corrections to the individual stars and implemented a weighted averaging approach for each segment of the FFD, where the weighting factor was determined by the number of stars with different detection thresholds contributing to that particular segment. For the young stars in the FUV flare sample, the results are plotted in Fig. \ref{fig:23} (the right panel), and as previously for individual FFD for AU Microscopii, show fitted broken power law at first slope is even shallower than 1.5 and show good agreement with the slope of 1.58 on the second part. While the 14 flares from AU Microscopii likely dominate the multi-star FFD distribution, it is worth noting that the other two stars, Karmn J07446+035 and 2MASS J02365171-5203036, known as active, each contribute 4 flares. Importantly, the additional flares in the sample are consistent in terms of ED and released energies, and more significantly, the fundamental parameters of these stars are somewhat similar. The found relation could hint at differences in energy release in FUV and at visible wavelengths, although for the young stars it seems that a broken power law with $\alpha_1$<1.5 and high energy flares at $\alpha_2$<2 can correctly represent the flare distribution in both wavelength ranges. We added the 6 active field stars to the distribution and plotted the results in Fig.~\ref{fig:45} in Appendix \ref{sec:appendix2}. These stars exhibit low to mid-size flares in FUV, and including them in the FFD, resulting in an even lower value in $\alpha_2$, and in minimal variation between $\alpha_1$ values.

\section{Discussion}
\label{sec:discussion}
\subsection{H$\alpha$ and rotation period as activity indicators}

Rotation period determination typically employs PDCSAP or SAP data analysed with the Lomb-Scargle method, as utilised in this study. This approach, while providing photometry for most of our sample, can encounter challenges when dealing with long rotation periods due to the limited observational windows of TESS. The constraints imposed by TESS's short-duration observations may lead to uncertainties in accurately identifying extended periodic signals, particularly for slowly rotating stars.
We do not report periods longer than 10 days, above the problematic periods of 13-14 days often detected in TESS stars. This is due to the TESS mission's observing strategy: it operates in a lunar-synchronous orbit with a period of 13.7 days, which subjects the telescope to background variations from reflected sunlight, resulting in periodic contamination that is challenging to remove. Therefore, the detection of periods longer than 12-15 days in TESS data presents a challenge (e.g., \citealt{2021AAS...23831407A,2020ApJS..250...20C,2020AAS...23527404H}).

In our analysis, we observed a relationship between H$\alpha$ emission, stellar mass, and age across our sample. We found that the equivalent width of the H$\alpha$ emission line generally increases with decreasing mass in young stars. This trend is evident in the H$\alpha$ versus (G - G$_{RP}$) diagram (see Fig.~\ref{fig:12}, right panel).
The majority of new rotation periods we reported are for stars belonging to the $\beta$ Pictoris and Tucana-Horologium YMGs, aligning with similar relations observed in previous studies. These findings are consistent with the analysis by \citet{2023ApJ...945..114P}, the rotation rate distribution alongside other youth indicators such as H$\alpha$ of the objects in Tucana-Horologium YMG, which is the second largest sub-group in our sample. They found that the equivalent width of the H$\alpha$ emission line for their sample increases with decreasing mass at the age of the group due to changes in photospheric contribution and not necessarily more magnetic activity \citep{2021AJ....161..277K}. In the H$\alpha$ against colour-colour diagram, the Tucana-Horologium YMG predominantly occupy a locus typically associated with chromospherically active stellar populations. \citet{2023ApJ...945..114P} conclude that stellar rotation period serves as a robust diagnostic for validating membership in young moving group associations. This conclusion is predicated on the relative ease of measuring rotation periods among young stellar populations.

YMG stars rarely rotate slower than $\sim$10 days. In our sample, we have only four stars belonging to the aforementioned groups with greater periods reported in the literature. A distinct transition in the stellar activity that was previously detected in the periods of rotation $\sim$10 days boundary. This transition was prominent in various observables, including flaring luminosities and amplitudes \citep{2016MNRAS.463.1844S,2019ApJS..243...28L}. The aforementioned boundary warrants consideration in investigations of field M dwarfs. Our analyses reveal also show the relationship between rotation period and age is clearly non-monotonic. Notably, the youngest stars in our sample exhibit rotation periods of approximately 4 days, while the older, from 24 to 50 Myr of age, groups include super-fast rotators. Our findings are in agreement with \citet{2022A&A...661A..29M} who found that the youngest stars in their sample rotate with periods of $\sim$ 4 days. \citet{2020A&A...638A..20M} also observed that the X-ray emission level of fast-rotating stars (i.e., those in the saturated regime) is not constant, suggesting the presence of rapid rotators even among older populations, which is in agreement with our results.

The diverse rotational behaviour observed in pre-main sequence stars underscores the complexity of angular momentum evolution in low-mass objects. Our youngest stellar population (4.5-5 Myr) exhibits inflated radii. As young stars contract, the conservation of angular momentum induces spin-up. However, disk-braking mechanisms can counteract this acceleration (\citealt{2010A&A...517A..88W}). This interplay between contraction-induced spin-up and disk interactions contributes to the intricate rotational evolution observed in young stellar populations, particularly in low-mass stars and brown dwarfs. In the 3-Myr-old Orion Nebula Cluster, \citet{2002A&A...396..513H} observed a correlation between stellar rotation rates and infrared excess emission. Specifically, slowly rotating stars exhibited significant infrared excess, indicative of the presence of inner gas disks. Conversely, rapid rotators, characterised by periods shorter than 3.14 days, displayed markedly reduced infrared emission. These findings are aligned with our results and lend support to the hypothesis that protoplanetary disks play a crucial role in modulating stellar angular momentum during the pre-main sequence phase of low-mass stellar evolution (see also e.g. \citealt{2009IAUS..258..363I,2015A&A...577A..98G,2012ApJ...746...43R}).

Another useful parameter for describing stellar magnetic activity-rotation relations is the Rossby number, which is depending on the convective turnover timescale and stellar mass. We calculated it for the stars in the sample from the empirical log $\tau$ following \citet{2018MNRAS.479.2351W}. The values of log~$\tau$ with errors up to $\pm$5\%  were reported for mass bins in the ranges of $\sim$ 0.04-0.19$M_{\odot}$. Therefore, it appears unfeasible not to use mass values found using MIST isochrones and the periods reported in this study. For stars in our sample, the found Rossby number estimates should be robust.

\subsection{The FFD analyses in mass-period bins}

In order to construct mass-period bins for our FFDs analyses, we calculated masses for the stars in the sample using TESS T magnitudes and MIST isochrones based on known ages for YMG. However, the isochrone fitting for age determination exhibits significant model-dependent variations. \citet{2015MNRAS.454..593B} demonstrated that applying different isochrone models to the $\beta$ Pic moving group's observed photometry resulted in age estimates spanning from 10 to 20 Myr. This phenomenon is not isolated to the $\beta$ Pic group; similar discrepancies have been observed in other nearby moving groups and open clusters \citep{2014MNRAS.445.2169M, 2023ApJ...954..134K, 2024MNRAS.528.4760L}. The disparities in age estimates derived from various isochrones can be attributed to differences in the underlying physics of stellar models, approaches to modelling interior mixing processes, and the stellar atmosphere models employed for bolometric corrections. Moreover, there are known colour discrepancies between observations and model predictions for certain photometry systems and evolutionary stages, which have been discussed in \citet{2025ApJ...979...92W}.
Since the stellar masses used in this work are primarily for the purpose of assigning stars into coarse bins to study the mass dependence of FFD behaviours, our isochrone-based masses fulfil the precision required. In this case, the faintest stars with uncertainty in the actual mass all fall in the mass bin with M$_\star$ < 0.11 M$_\odot$ as the approximate upper value for the mass of a star observed with T magnitude larger than the largest available one in the MIST isochrone tables for all age bins for stars in our sample. Given the inherent complexities and potential for introducing additional uncertainties, we have elected to forgo further efforts to refine the mass determinations.

Prevailing models of stellar structure and evolution suggest that stars below $\sim$0.28-0.35 solar masses maintain fully convective interiors throughout their lifetimes (e.g., \citealt{2018A&A...619A.177B,1997A&A...327.1039C,2000ARA&A..38..337C}). The two upper rows in the grid shown in Fig.~\ref{fig:23},  are likely consisting entirely of fully convective stars and provide a visual guide that aligns with the physical properties of the stars, enhancing the interpretability of our mass- and rotation-period-based binning strategy.

The lowest-mass bin for stars with M$_\star$ < 0.11M$_\odot$ shows the tendency to obey $\alpha_2$ $\sim$ 2.0 and steeper in the high-energy tail, while the more massive stars FFDs depart to lower values of $\alpha$ even at the larger energy cut-offs or larger ED. In several previous studies \citep{2019ApJ...881....9H,2019A&A...622A.133I, 2021A&A...645A..42I}, the large $\alpha \sim$ 2.0 coefficients were spotted for stars of different ages given that the authors use high energy cut-off for fitting power laws or subsequent analyses as in \citet{2020ApJ...905..107M}. Analysing the FFDs in young stars in a wide variety of spectral types, \citet{2024AJ....168...60F} found that in their sample, the young stars obey $\alpha$=1.5. Our results show that only the fast-rotating/very low-mass stars in our sample, whatever the age, show either a quick switch to $\alpha_2~\sim$2.0 regime or the massive high-energy tail sometimes even steeper than that. However, for larger masses or rotation periods, the population tends towards a power law coefficient $\alpha_2~\simeq$1.5-1.7, until stars with P$_\mathrm{rot} \geq$ 30 days again exhibit a tendency to follow a power law with larger $\alpha_2$. Notably, all individual stellar FFDs with sufficient flare counts exhibit a distinct break in their power-law slopes (see e.g. Fig.\ref{fig:23}, left panel for AU Microscopii). This observation naturally leads to the idea of a piecewise power law, which was not previously fully implemented in the studies of FFDs in young and active M dwarfs.

\subsection{The FFDs in young dwarfs are governed by piecewise power law}

The observed FFDs in mass-period bins deviate from a single power law at lower energies, which cannot be fully explained by detection limitations. While distance can affect flare detection in general (i.e. for the distant stars the detection will be only available for the most energetic events), for stars within 200 pc which is primary zone for TESS survey, other factors such as stellar type, intrinsic flare energy, and observational cadence are likely to have a more significant impact on the detectable flare distribution than distance alone. \citet{2014ApJ...797..121H} demonstrated that even flares with durations of 10 minutes (corresponding to amplitude 0.002 and log E$_{Kp}$ $\sim$ 30.5) are readily identifiable in Kepler data with 1 minute cadence, suggesting that a significant number of such flares is likely being detected, rather than going unnoticed. TESS's overall sensitivity to weak flares remains somewhat lower than that of Kepler, despite this enhanced temporal resolution \citep{2020ApJ...890...46T,2019MNRAS.489..437D}. Nevertheless, the inclusion of fast-cadence (20 s) TESS light curves for numerous stars in our sample, combined with the injection-recovery routine (see Appendix \ref{subsec:absdetect} for details), significantly enhanced our ability to reliably detect and validate smaller flare events.

The deviation from a single power law may indicate that lower energy flares follow a different power law distribution compared to high energy flares, as discussed by \citet{2011SSRv..159..263H} for solar flares.  \citet{2010ApJ...710.1324W} presented evidence suggesting a deviation from the power-law distribution in flare sizes for a small solar active region and suggested a piecewise Poisson distribution, correlating with the evolving magnetic topology of the region under observation. \citet{2018ApJ...858...55P} found deviations from a single power-law dependence for some targets in their small sample of ultra-cool dwarfs. Studying active and inactive M dwarfs in Kepler data, \citet{2014ApJ...797..121H} conducted analyses that indicate that the power-law fit does not persist at lower energies, even though they lie above the detection limit. They surmised that this discrepancy may arise from either the incorporation of low-energy flares into complex events or a genuine turnover in the power-law distribution.

Observed flare events may actually be composite phenomena comprising multiple concurrent flares of varying energies. The apparent break in the power law distribution may stem from the challenge of distinguishing overlapping low-energy flares, which are often classified as single, more energetic events. This issue is exacerbated by the lack of spatial resolution in Kepler and TESS data. While resolving these complex flare structures might yield a distribution closer to a simple power-law, our current observational constraints necessitate the use of a broken power-law model, as it may more accurately represent the true flare energy distribution, accounting for both physical differences in flare generation mechanisms and observational limitations.   This approach provides a more realistic representation of stellar activity, crucial for studies of exoplanet atmospheres.

\subsection{Activity patterns}

The observed patterns in FFDs within our parameter space probe the intrinsic relationships among stellar mass, rotation period, and various activity indicators. These indicators include the occurrence rates of small-scale, moderate, and extreme flaring events. The broken power law approach will allow us to accurately describe small-scale and mid-size flares, which is applicable in flare models to study the radiative environment for the planetary systems of M dwarfs. As for the super-flare events, they require special consideration, as our analysis indicates that such events are exceedingly rare — sometimes even rarer than suggested by a power law with $\alpha \sim 2.0$ — regardless of the age, mass, or rotation period observed in M dwarfs, and even for the youngest and most active stars these events' frequency will fall short of the slope of $\alpha \sim 1.5$.

As reported in \citet{2023ApJ...945..114P} in their evaluation of activity patterns for M dwarfs belonging to Tucana-Horologium, for 0.6 < (G - G$_{RP}$) < 1.2 , the H$\alpha$ values are independent of rotation period. The Rossby number, which combines rotation period and convective turnover time, is a key indicator of stellar magnetic activity, and its activity relation evolves with stellar age.
It was therefore of interest to investigate whether the H$\alpha$ activity indicator could signal changes in the FFDs of young stars. In Fig.~\ref{fig:46}, we plotted the logarithm of the Rossby number for our mass-period bins in the sample against the H$\alpha$ equivalent width (EW H$\alpha$). The relationship between age and EW H$\alpha$ is apparent, with the largest values of EW H$\alpha$ corresponding to the youngest stars. However, the previously observed nuanced relationship between rotation periods and FFDs behaviour does not seem to manifest itself distinctly in the EW H$\alpha$ measurements. Overall, we observe a large variation in the behaviour of the stars' activity, even for stars with nominally similar properties.

\subsection{Comparing scalar flare rate across stellar properties}
Since observable flares occur with a wide range of energies for a given star, such a frequency can be seen as for flares of a given energy or larger (e.g., 10 flares per year with E $\geq10^{32}$ erg). This is equivalent to evaluating the FFD at a given energy. The specific flare rate has been used in some capacity for many years for comparing flare activity levels between stars and was noted by  \citet{2016ApJ...829...23D} as an effective metric for comparing flare activity levels between stars at different distances. \citet{2020ApJ...905..107M} selected a threshold flare energy of 3.16 $\times$ 10$^{31}$ ergs ($\log_{10}(E)=31.5$) in the TESS bandpass for their sample of low mass stars at distances less than 15 pc. This threshold was chosen by the authors because it corresponds to the energy at which the completeness function reaches or exceeds 50\% for all stars in their sample of M and L dwarfs and ensured a consistent level of flare detection completeness across our entire stellar sample in the TESS observations.

The YMGs to which the stars in our sample of M-K dwarfs belong are at mean distances from 29 to 172 pc and brighter than TESS magnitude 14. We determined that the mean energy of the photometric flares in our sample is 1.9$\times$ 10$^{32}$ ergs. This average energy is higher than the aforementioned threshold and could potentially be attributed to the heightened activity levels characteristic of young stars and aligns with the expectation that younger stellar populations exhibit more energetic flaring events, reflecting their enhanced magnetic activity. It was established that rapidly rotating stars tend to show more flares, but the highest flare rates are not necessarily found among the fastest rotators \citep{2020A&A...637A..22R}. And the youngest population of stars in our sample indeed exhibit energetically highest flares while their rotation periods put them in P$_{rot}\in$[4,7] days bin.

\begin{figure*}\resizebox{\hsize}{!}{
   \centering
   \includegraphics{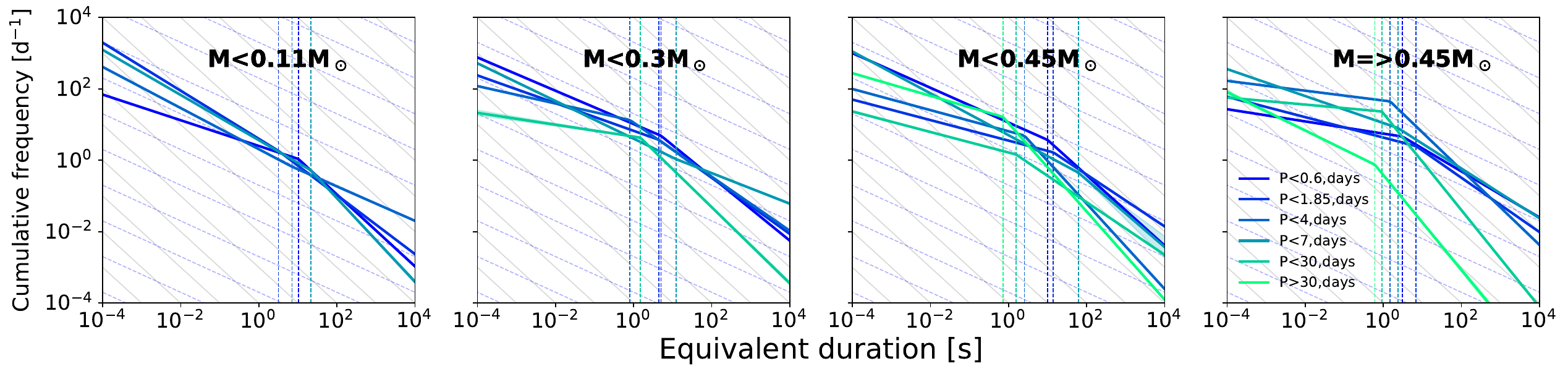}
    }
\caption{The broken power law relation for FFDs found in this study sample of young and field stars is plotted for four mass bins. The colour of the lines indicates the period bins. The lines represent the broken power law slopes $\alpha_1$ and  $\alpha_2$ for successful fits. Uncertainties are plotted as light envelopes in the same colour as period bins.}
\label{fig:47}
\end{figure*}

In Fig.~\ref{fig:47} we plotted the fitted relations for four mass bins used in our study, with coloured lines representing different period bins. We propose that these relations can be applied in modelling the distribution and evolution of flaring activity of young stars with corresponding stellar properties, in conjunction with their rotation histories.
This approach would enable more precise assessments of the radiative environment in evolutionary models of planet-star interactions. Such refinement could significantly enhance our understanding of how stellar activity impacts planetary evolution in young stellar systems.

\subsection{Evaluating UV FFD behaviour in YMG members}

In this study, we utilised archival FUV observations of low-mass stars to evaluate the UV predictions of existing FFDs for photometry observations. We employed HST TIME-TAG spectroscopy to produce time-resolved light curves and further implement our flare finding and injection-recovery routines. In Fig.~\ref{fig:23} we presented the resulting FFD for one star, AU Microscopii, in optical range (left panel) and FUV (centre panel), along with the multi-star sample FFD in UV (right panel). The UV flare distribution demonstrates a broken power law with $\alpha_2$=1.48, while the TESS observations show much steeper $\alpha_2$=1.8. It could indicate the different flare frequency behaviour in the different wavelength ranges. However, \citet{2023ApJ...951...33T}, studying the multi-wavelength activity of AU Microscopii in NUV photometry and soft X-ray, found that both wavelength range distributions show similarity in $\alpha \sim$1.72,  slightly shallower compared to our results in TESS bandpass for the same star. For FUV activity in AU Microscopii, we found a much shallower second slope. The discrepancies may be attributed to the incompleteness of our sample or variations in the FFD behaviour across different regions of the spectrum.

Studying NUV flare energies and frequencies, \citet{2024ApJ...971...24P} and \citet{2023ApJ...955...24R} found that the flares in this range are much more frequent in comparison to optical. For the FUV range, \citet{2024MNRAS.533.1894J} reports a higher frequency of ultraviolet flares exhibiting minimal or undetectable corresponding optical signatures. Analysis of HST COS FUV data revealed power-law coefficients of 1.58 for flare energies and 1.61 for EDs in M-type stars belonging to the Tucana-Horologium YMG (\citealt[Fig.~1]{2020RNAAS...4..119L}). Although the authors' analysis is based on a limited sample of 18 flares, the $\alpha$ values they report are consistent with our results when considering $\alpha_2$ alone, as identifying a broken power-law becomes difficult with small sample sizes. For the field stars, they found enhanced $\alpha$ coefficients of 1.74 and 1.77 for energy and ED, respectively. However, incorporating older stars in our multi-star FFD analysis (Fig. \ref{fig:45}, Appendix \ref{sec:appendix2}) yielded an even lower $\alpha_2$=1.39, for which our larger sample size could be the explanation. Our results suggest that flares with longer EDs are more frequently detected in the FUV compared to the visible wavelengths for a given star, and the same tendency is apparent in the multi-star analyses (Fig.~\ref{fig:23}, right panel). This disparity warrants further investigation into the underlying physics of stellar flare generation across different spectral ranges and implies that the processes driving flare occurrence may differ between FUV and visible emissions, particularly in young stellar objects.

\citet{2020RNAAS...4..119L} detected FUV flares from the optically quiet stars. \citet{2024MNRAS.533.1894J} found that various models (e.g. 9000 K blackbody, the combination of it and the Great AD Leo flare characterisation by \citet{1991ApJ...378..725H}, and Fiducial flare model proposed by \citet{2018ApJ...867...71L} significantly underestimate emission across all ages, masses, and activity levels, with discrepancies reaching up to four orders of magnitude for combined FUV continuum and line emission. The authors noted that underestimation is even greater for individual emission lines, highlighting the need for improved modelling approaches to accurately represent stellar emission processes. This suggests that the spectral energy distribution in flares varies between the partially convective and fully convective interiors. This could suggest that the transition to a fully convective interior alters the underlying magnetic field generation mechanisms, leading to variations in energy fractionation across all wavelengths. \citet{2024MNRAS.533.1894J} also worked with the Tucana-Horologium YMG to investigate potential age-dependent changes in flare model accuracy. By comparing UV correction factors between the 40 Myr sample and field age stars within the same mass range, they identified that models underestimate FUV emission less severely as stars age.

Our sample of the UV flares is small and relatively incomplete compared to the extensive photometric flare dataset, which enabled more in-depth analyses. However, in this small sample, we spotted the power law for the second slope of $\alpha_2$=1.58, which is similar to found by \citet{2022AJ....164..110F} for the stars belonging to YMG ($\alpha \sim 1.5-1.6$). Concurrently, the entire sample of young and field stars of UV flares, examined in this study and plotted in Fig.~\ref{fig:45} in Appendix \ref{sec:appendix2}, displays shallow slopes. While our UV sample size is not sufficient for definitive conclusions, the observed discrepancy in flare frequency behaviour for AU Microscopii and presumably for many young stars as seen in Fig.~\ref{fig:23} suggests potentially distinct flare production in the UV and visible wavelength regimes and shows some support to the broken power law model for flare energy distributions. We note that \citet[Fig.10]{2023ApJ...951...33T} demonstrates that the FFDs of AU Microscopii, along with those of the reference stars YY Gem and Proxima Centauri, may plausibly be characterised by a piecewise power-law model.

The FUV regime enables detection of lower-energy flares through direct observation of chromospheric activity, whereas optical wavelengths predominantly reveal high-energy white-light events requiring photospheric heating. To be detectable in the TESS bandpass, flares must inject sufficient energy to heat the photosphere, resulting in reduced observational sensitivity to low-energy white-light events. This is supported by multiple detections of UV flares with non-detectable or weakly detectable optical counterparts \citep{2024MNRAS.533.1894J,2023ApJ...944....5B,2024ApJ...971...24P,2023ApJ...955...24R}. Consequently, the FUV regime exhibits enhanced detectability of lower-energy flare events compared to optical observations. The alignment between the FUV's $\alpha_2$ slope and the low-energy optical regime's $\alpha_1$ slope likely arises because FUV observations probe the low-energy FFD, and the lack of high-energy events in our sample of UV flares reflects the limited observational baselines of COS/HST data (typically hours per target compared to TESS's 27-day continuous monitoring per sector). Future studies incorporating additional simultaneous optical, NUV, FUV and X-ray observations will enable better constraints on the relative occurrence of flares across these wavelength regimes for low-mass stars.

\section{Conclusions}
\label{sec:conclusions}

Our investigation into flare activity in young M dwarfs, particularly those in young moving groups, has yielded crucial insights into their stellar flare behaviour with the potential impact on atmospheres of exoplanets around them. We demonstrate that stellar rotation period and mass serve as more fundamental parameters governing flare frequency distributions than age alone, suggesting more complex mechanisms govern frequency-flare distributions than previously hypothesised.

We propose a novel approach using a piece-wise power law model for low-to-mid-size and large flares in young M dwarfs, allowing for a more nuanced representation of flare behaviour across energy ranges. This method, combined with known rotation periods, enables the generation of more realistic flare event sequences, crucial for assessing exoplanet atmosphere stability and potential habitability. While the conflation of large and small flares does not significantly impact the total energy budget, it can lead to atypical power-law relationships. Observational biases at distribution extremes must be considered when interpreting stellar activity patterns and their implications for exoplanetary environments. Our broken power law approach in FFDs can improve the modelling of activity patterns in specific young M dwarfs.

These findings have significant implications for future exoplanet characterisation missions, particularly those studying M dwarf terrestrial planet atmospheres. By providing a more accurate picture of the flare environment, we can better predict challenges in detecting biosignatures and assessing habitability.

\section*{Data Availability}
The dataset comprising stellar parameters obtained in this study is available at the CDS via anonymous ftp to cdsarc.u-strasbg.fr (130.79.128.5) or via the CDS website (https://cdsarc.cds.unistra.fr/) under catalogue J/A+A/XXX/XXX (doi:xxx).

\begin{acknowledgements}
We acknowledge financial support from the Research Council of Norway (RCN), through its Centres of Excellence funding scheme, projects number 332523 (PHAB, Centre for Planetary Habitability) and number 262622 (ROCS, Rosseland Centre for Solar Physics). We thank Konstantin Herbst (PHAB), Alexey Pankine (Space Science Institute, USA) and Adalyn Gibson (University of Colorado Boulder, USA) for the fruitful discussion. We thank the anonymous reviewer for the thoughtful feedback that helped to improve this manuscript.
\end{acknowledgements}

\bibliography{bibliography}
\newpage
\begin{appendix}
\section{Detailed Methodology}
\label{sec:appendix1}
\subsection{H$\alpha$ emission data and analyses}

The intrinsic faintness of M dwarfs has made the H$\alpha$ line the primary diagnostic spectral feature of chromospheric activity. For stars in the sample, we retrieved H$\alpha$ emission detection, using the literature (specifically, the EW H$_\alpha$ values come from \citet{2014AJ....147..146K} for Tucana-Horologium YMG, from \citet{2015A&A...575A...4F} for Chamaeleon I YMG and from \citet{2019AJ....157..234S} for $\beta$ Pictoris YMG) and the Gaia GAIA DR3 dataset. From the latter, we obtained a H$\alpha$ pseudo-equivalent width (pEW) computed as the integrated normalised flux over a wavelength domain (646-670 nm, i.e. at both sides of 656.5 nm). Measuring the H$\alpha$ emission line presents significant challenges, particularly for stars with temperatures below 4000 K. These difficulties arise from two main factors: the limited resolving power of BP and RP spectra, and the sharp decrease in transmission efficiency at the H$\alpha$ wavelength, and the issue is thoroughly discussed in \citet{2023A&A...674A..28F}. To address the spectral complexity in cooler stars and mitigate blending effects with non-hydrogen species, the authors implemented a correction strategy. They subtracted the equivalent width derived from the nearest synthetic spectrum below 5000 K within the GSP-Phot atmospheric parameter space (denoted as $\mathrm{EW,H\alpha},model$).

\begin{equation}\label{eq:11}
\mathrm{EW,H\alpha}  =
\begin{cases}
\mathrm{pEW,H\alpha} & \text{if } T_\mathrm{eff} > 5000 \text{ K} \\
\mathrm{pEW,H\alpha} - \mathrm{EW,H\alpha,model} & \text{if } T_\mathrm{eff} \leq 5000 \text{ K}
\end{cases}
\end{equation}

We proceed to compare the H$\alpha$ pseudo-equivalent width measured by ESP-ELS (Extended Stellar Parametrizer for Emission Line Stars, a module within the Gaia astrophysical parameters inference system (Apsis) used in Gaia DR3, \citealt{2023A&A...674A..26C}) to the published equivalent width values. In Fig.~\ref{fig:24} in Appendix \ref{sec:appendix1}, left panel, we plot the age of young stars in our sample against EW H$\alpha$, both including and excluding $\mathrm{EW,H\alpha, model}$ (note that our sample does not contain stars with $T_\mathrm{eff} > 5000$ K). In the same figure, in the left panel, we plotted the relation for EW H$\alpha$ reported in the literature and EW H$\alpha$ from ESP-ELS. Including and excluding the model would not lead to the agreement with literature, while including the model results in underestimating the H$\alpha$ strength, and excluding it leads to overestimating EW H$\alpha$ for the stars of the age 24-800 Myr. For very young stars, represented in our sample by two groups of 4.5-5 Myr, the data in the literature is scarce. Despite the Gaia pEW H$\alpha$ values for the youngest stars showing disagreement with the literature in some cases, we consider pEW H$\alpha$ from Gaia in the form recommended by Eq. \ref{eq:11} as the estimate of the H$\alpha$ line strength for the youngest stars in our sample due the absence of these EW H$\alpha$ values in literature.

\begin{figure*}[h!]
\sidecaption
\includegraphics[width=12cm]{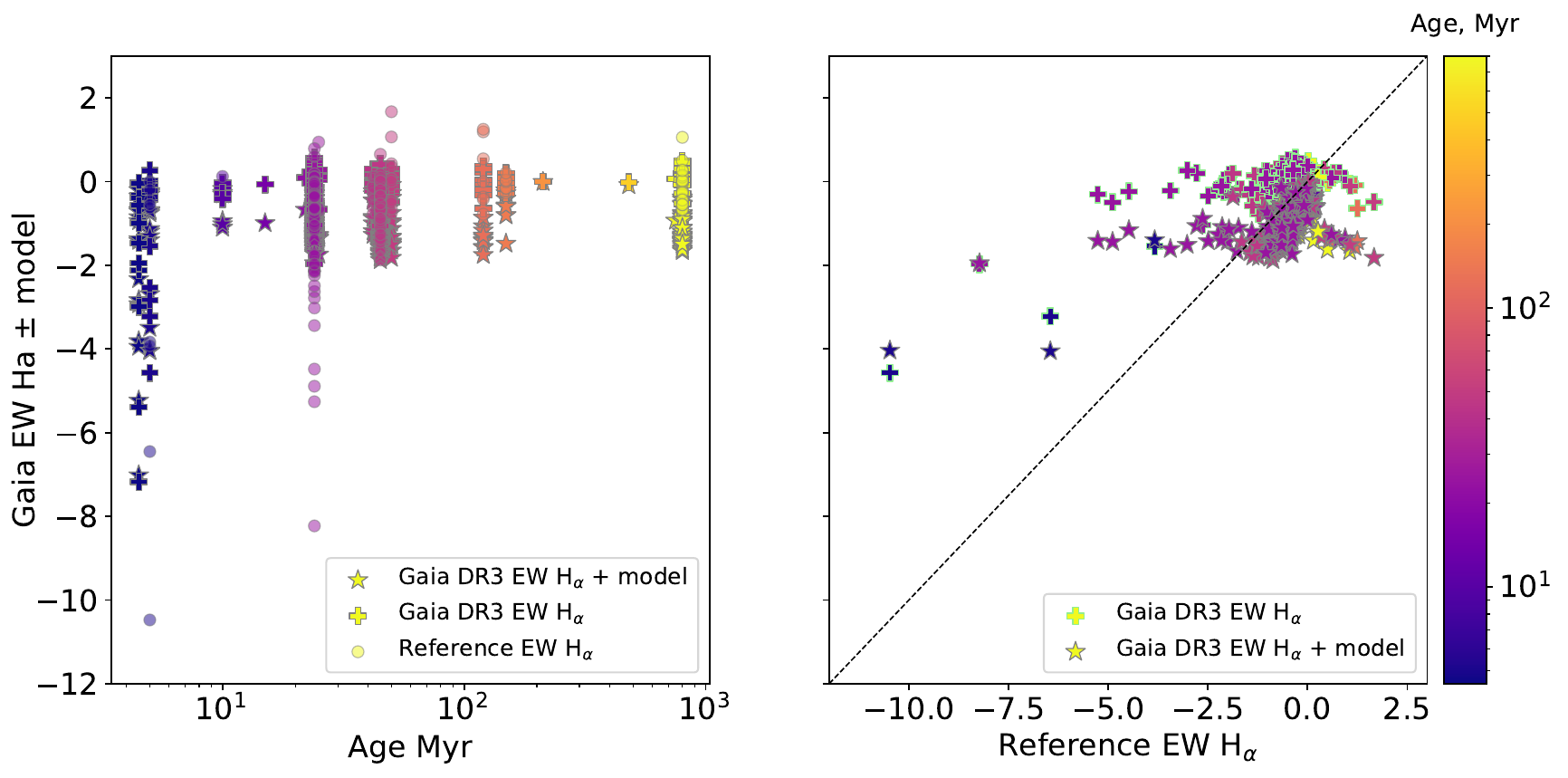}
\caption{Comparison of EW H$_\alpha$ from Gaia DR3 and literature. Left panel: For the sample of stars, three values of equivalent width of Balmer $\alpha$ line were plotted according to eq.\ref{eq:11}: Gaia DR3 EW H$_\alpha$, Gaia DR3 EW H$_\alpha$ with added model and EW H$_\alpha$ found in literature. Right panel: the diagram presents the effect of the model from the relation described in Eq.\ref{eq:11}.  Ages of the sample's stars are colour-coded from youngest (blue) to oldest (yellow).}
\label{fig:24}
\end{figure*}

For analyses of the activity of stars in our sample, we used the Rossby number, the indicator that connects such parameters as mass and period of rotation. The values of log $\tau$ were reported for mass bins with errors up to $\pm$5\% of the value. Therefore, it seems impractical not to use mass values found using MIST isochrones and the found periods because most of the sample stars showed rotation periods within error intervals of the values reported previously or did not differ from them more than 10\%. The uncertainties in $M_{\star}$ and $P_{\mathrm{rot}}$ propagate into our Rossby number calculations. We utilise $\log \tau$ values from \citet{2018MNRAS.479.2351W}, reported with $\pm$5\% uncertainty, which aligns well with our mass bins derived from MIST isochrone-based $M_{\star}$ estimates. The additional source of uncertainty in Ro stems from $P_{\mathrm{rot}}$. For most stars, our derived $P_{\mathrm{rot}}$ values align well with previous reports, typically well within 10\%.  Consequently, we estimate the typical uncertainty in Ro to be approximately 10\%, primarily driven by the combined uncertainties in $\tau$ and $P_{\mathrm{rot}}$ and does not compromise the reliability of the calculated Rossby number estimates.

\subsection{Light curve data processing}

The stars' sample analysed in this study was observed by both the Kepler and TESS missions. For Kepler light curve data we used the Kepler Science Processing Pipeline \citep{2010DPS....42.2709J}, which processes science data collected from the Kepler Photometer: raw and systematic error corrected flux time series. For K2 data, we used K2 pipeline \citep{2016ksci.rept....9T}produced light curve files by using simple aperture photometry. For TESS we used the TESS Science Processing Operations Center (SPOC) pipeline \citep{2020RNAAS...4..201C}, which generates and calibrates target light curves.

From the Mikulski Archive for Space Telescopes (MAST)\footnote{http://archive.stsci.edu}, we downloaded processed data from both long cadence (LC), 1800 s, and short cadence (SC), 60 s, star light curves from Kepler (\citealt{2010DPS....42.2709J}), LC 1800 s and SC 60 s cadence data from K2 \citep{2014PASP..126..398H}, following the LC (1800 s/600 s/200 s) and SC data (120 s/ 20 s) from TESS missions \citep{2016SPIE.9913E..3EJ}. We downloaded light curves, corrected the detection efficiency, and proceeded to de-trend and remove the variability. We used the light curves first for the period of rotation determination. For several stars, we reported new periods, in other cases, we used values we found if they were within $\pm$10\% of those found in literature, and if we were unable to find or determine the rotation period, the star was excluded from the sample.

Utilizing the light curve data described above, we derived stellar rotation periods from photometric modulations caused by rotating starspots. These variations typically occur on longer timescales compared to flare events. For stars in the sample, we started with downloading LC TESS data, which is available at 1800 s intervals for TESS Cycles 1 and 2 (sectors 1-26), the 600 s cadence data in Cycles 3 and 4 (sectors 27-55), and the 200 sec data in Cycles 5 and 6 (sectors 56-83). We continue by downloading all available 90-day LC quarters and sectors of Kepler and K2 for our sample. Kepler has been observing stars at 29.4-minute intervals as LC targets for the primary purpose of detecting transiting planets. For determination of the rotation period and hereafter we used Lightkurve module \citep{2018ascl.soft12013L}.

\subsection{Flare detection}
\label{subsec:absdetect}

In order to be able to work simultaneously with light curves from TESS and Kepler and collected data from UV instrument COS onboard of HST, we implemented a Python routine "YMDF" based on the Python module `altaipony` \citep{2021A&A...645A..42I,2022ascl.soft01009I}, which was designed for finding and analysing flares in Kepler, K2 and TESS photometry. The `altaipony` module uses the Automated flare finding routine Appaloosa, the open-source flare finding and analysis procedure written in Python by \citet{2016ApJ...829...23D} for Kepler and the algorithm, designed by \citet{2016MNRAS.459.2408A} for de-trending and variability removal, hereafter `autofinder`. The module also uses `lightKurve` Python package for downloading the data. For the de-trending of data, we apply a Savitky-Golay filter \citep{1964AnaCh..36.1627S} to the light curve. The filter is based on the method of data smoothing by local least-squares polynomial approximation, and it works by fitting a polynomial to a small window of adjacent data points and using this polynomial to estimate the value of the central point in the window. The window is then shifted along the signal, and the process is repeated for each point in the signal. The de-trending of the light curve allows to flag outliers, and they were clipped at 3$\sigma$ iteratively. This method was chosen by \citet{2021A&A...645A..42I} because it is sufficiently quick and provides satisfying results in both Kepler and TESS light curves. At the next step, for the detection of flare events, we used the following conditions: the flare candidate definition follows the criteria in \citet{2015ApJ...814...35C}:

\begin{equation}\label{eq.1}
   x_i - \bar{x_L} < 0
\end{equation}

\begin{equation}\label{eq.2}
\centering
   \frac{|x_i - \bar{x_L}|}{ \sigma_L} \geq N_1
\end{equation}

\begin{equation}\label{eq.3}
 \frac{|x_i - \bar{x_L}| - w_i}{ \sigma_L} > N_2
\end{equation}

\begin{equation}\label{eq.4}
ConM \geq N_3,
\end{equation}
where $x_i$ is the current flux value, $\bar{x_L}$ and $\sigma_L$ are the local mean standard deviation for a given light curve, $w_i$ is the photometric error at epoch $i$ , and $ConM$ is the number of consecutive points which satisfy Eq.\ref{eq.1},~\ref{eq.2},~\ref{eq.3}. The values of N$_1$, N$_2$, N$_3$ are taken to be 3, 3, and 2, respectively (the recommended values according to \citealt{2015ApJ...814...35C,2022ascl.soft01009I}). This model simplifies the task of detecting significant statistical outliers and can identify flare candidates with 3$\sigma$ above the iterative median flux, regardless of intrinsic variability.

There are several effects the flare finding algorithm can produce: large flares may appear larger, and small flare candidates may be unaccounted for. The former occurs when stellar variability adds flux to the flare, which was mitigated by de-trending preprocessing. The latter arises when small flares in the de-trended light curve fail to meet the criteria distinguishing them from quiescent flux. To account for these effects, we applied an injection-recovery routine, comparing recovered events to injected ones. Following \citet{2014ApJ...797..122D}, who suggest a piecewise model for flare shapes, our module produced and injected an array of artificial flares constructed following the flare model in \citet{2022AJ....164...17T} into every light curve where at least one flare event was observed.

The flare recovery completeness has also been computed throughout each light curve, where we found the flare activity using artificial flare injection tests, and we used the tools and the empirical flare model, first suggested by \citet{2016ApJ...829...23D}  and implemented in the YMDF module. The artificial events were introduced to each light curve at randomly generated times with varying amplitude and duration while avoiding overlap with real flare signatures. A flare was then considered recovered if the flare peak time was contained within the start and end times of any resulting flare event candidate. After relating all successful and failed detections to each other, the recovery rate as a function of equivalent duration was returned. The method for calculating the probability of recovering tiles up the sample of artificial flares into amplitude and duration bins twice: first, it tiles up the sample into a matrix based on the recovered amplitude and durations, second, it does the same with the injected flare parameters, and so include also those injected flares that were not recovered. The first matrix is used to map each flare candidate’s recovered equivalent duration to a value that accounts for losses dealt to the equivalent duration by photometric noise and introduced by the de-trending procedure. The typical injected amplitude and duration of flares in that tile of the matrix can then be used by the second matrix to derive the candidate’s recovery probability from the ratio of lost to recovered injected flares.

\subsection{White-light flare analysis}
\label{sec:appendix_flare}

Using formalism in Sect.\ref{subsec:wlca} (Eq.\ref{eq:5}, \ref{eq:6}, \ref{eq:61}, \ref{eq:7}), we calculated the energy of every flare exhibited by stars in our sample. For these calculations, we obtained the Kepler Instrument Response Function for low resolution from the Kepler Instrument Handbook \citep{2016ksci.rept....1V}, and the
TESS response function\footnote{https://heasarc.gsfc.nasa.gov/docs/tess/data/}. We note that by this approximation we miss at least the $\sim$27\% of the continuum flare flux that resides in the U band relative to the total in UBVR \citep{1992ApJS...78..565H} and flux from emission lines that lie outside the Kepler band \citep{2013ApJS..207...15K}. As a consequence, E$_{Kp,\mathrm{flare}}$ should be considered a lower limit to the total released energy of the flare \citep{2019A&A...622A.133I}.

\citet{2017ApJ...841..124V} showed that the flare energies from Kepler can be converted to flare energies in the TESS bandpass as follows:

\begin{equation}\label{eq.10}
E_{\mathrm{TESS}} = (0.72 \pm 0.02)E_{\mathrm{Kepler}}.
\end{equation}

This results in a 0.14 shift on the logarithmic energy scale. While working with our sample from both Kepler and TESS missions, we recalculated found values of energy in the TESS bandpass, as the majority of our sample stars' light curves originated from TESS survey.

The 'mcmc' model was considered preferable by the aforementioned authors, and they follow \citet[Eq. 24]{2004ApJ...609.1134W} using the Markov chain Monte Carlo method (MCMC), and sampling from the joint posterior distribution with a constant prior. For quicker results, we used Modified Maximum Likelihood Estimator (MMLE), which is detailed in \citet{2009MNRAS.395..931M}, to find the slope $\alpha$. The least squares method was used to fit to estimate the intercept $\beta$. Having established a common value of $\alpha$, we can fix this value to estimate the flare rate for stars with fewer than 5 flares, and to obtain a more precise value of the flare rate for stars with 5 or more flares. This simplified approach is sufficient as the resulting values only serve as an initial estimate for assessing the power law distributions in the flare sample. We proceed with these calculations for all stars in the sample.

In the previous studies \citep{2019ApJ...881....9H,2015ApJ...814...35C,2020ApJ...905..107M,2013ApJ...762...41G,2020A&A...637A..22R}, the ED, energy or duration cut-off is usually introduced for power law coefficients calculation. We quantified the completeness in our study of FFD: to assess the effectiveness of the aforementioned 'autofinder' algorithm, we conducted a completeness study. The corrections to the number and energy of detected flare events are based on two factors: the detection probability and the energy recovery ratio. For every flare in the sample, we retrieved the recovery probability P$_{rec}$ and used only flares with P$_{rec}$ >0.25 instead of cutting off the flares that can not be considered super-flares. We did that because we are interested also in the middle-range energy flares, as they comprise the majority of young M dwarfs' flares and it can potentially largely affect exoplanets’ atmospheres.

\subsection{HST/COS flare analyses}
We retrieved archival data from the HST COS G130M grating, identified by program IDs: 14784 (PI Shkolnik); 15071, 16033 (PI Froning); 16482 (PI Roman-Duval); 15955 (PI Richey-Yowell); 16164 (PI Cauley); 17428 (PI France); 11533 (PI Green). This grating is centered at 122.2 nm, with resolving power $R=\lambda/\Delta\lambda\sim$12 000-16 000, following the configuration used in \citet{2019ApJ...871L..26F,2022AJ....164..110F}. This setup offers spectral coverage spanning approximately 106.0–136.0 nm, with a detector gap between 121.0–122.5 nm. This gap strategically masks the intense Ly$\alpha$ emission feature to avoid detector saturation. The same COS setting was used for all analyses. We selected this specific grating based on two key considerations. Firstly, it provides coverage of chromospheric and transition region emission lines, which are crucial for investigating far-ultraviolet (FUV) emission and activity in low-mass stars. Secondly, the abundance of observations available using this grating enables us to maintain self-consistency throughout our study, enhancing the reliability and comparability of our results across multiple targets.

For TIME-TAG mode observations, CALCOS creates a corrected events list (i. e. "corrtag" file) that preserves the time-stamped photon events. To create time-resolved flux data from COS observations using the G130M grating, we extracted spectra from the corrtag file. Each extraction involved selecting events within the specified time range, applying wavelength calibration, and converting counts to flux using the appropriate sensitivity curve. For the data reduction, we binned the observations into 20 s exposures to balance high temporal cadence with a sufficiently high signal-to-noise ratio (S/N) per bin, as previous studies (i.e. \citealt{2022ApJ...926..204H}) suggest that the time resolution can impact measured flare amplitudes and energies.

The initial AltaiPony package, on which we are based our "YMDF" routine, was made to work with high quality, uniformly processed final data from these TESS and Kepler surveys (i.e. PDS-SAP fluxes), that is, with already removed instrumental and astrophysical variability for the signal.  In the case of COS-HST data reduction, the CALCOS pipeline, as previously described, corrects the data for instrumental effects, including noise, thermal drifts, geometric distortions, and pixel-to-pixel sensitivity variations, generates an exposure-specific wavelength-calibrated scale, and extracts and produces the final one-dimensional, flux-calibrated, time-resolved spectrum. We, therefore, implemented the same framework in the YMDF, including routines: i) de-trending flux; ii) find, characterise, inject and recover flare events; iii) calculating corrected EDs (in seconds) and recovery probability; to work with time-resolved flux integrated across FUVB-FUVA range for 20 s exposure time steps.

\section{Supplementary tables and figures}
\label{sec:appendix2}

\begin{table}
\scriptsize
\caption{\label{tab:12}Broken power law coefficient for the mass-period binned sample of young and field stars}
\centering
\resizebox{0.45\textwidth}{!}{\begin{tabularx}{9cm}{l | c c c c} \hline
\multicolumn{1}{l}{}&\multicolumn{4}{c}{M$_\star$<0.11M$_\odot$}\TBstrut\\
\hline
Period range (days) & $\alpha_{1~ini}$ & $\alpha_1$ & $\alpha_2$ & x$_{break}$   \TBstrut\\
 \hline
P$_{rot}$<0.6      & 1.459 & 1.360$\pm$0.003 & 2.007$\pm$0.004 & 10.333$\pm$0.055 \TBstrut\\
0.6<P$_{rot}$<1.85 & 1.644 & 1.673$\pm$0.006 & 1.827$\pm$0.004 & 3.197$\pm$0.101 \TBstrut\\
1.85<P$_{rot}$<4.0 & 1.500 & 1.58$^1$ & 1.48$^1$ & 7.11$^1$ \TBstrut\\
4.0<P$_{rot}$<7.0  & 1.500 & 1.633$\pm$0.002 & 2.171$\pm$0.018 & 21.365$\pm$0.381 \TBstrut\\
\hline
\multicolumn{1}{l}{}&\multicolumn{4}{c}{\ 0.11$\leq$M$_\star$<0.3M$_\odot$} \TBstrut\\
 \hline
P$_{rot}$<0.6      & 1.491 & 1.466$\pm$0.002 & 1.891$\pm$0.003 & 5.016$\pm$0.032 \TBstrut\\
0.6<P$_{rot}$<1.85 & 1.435 & 1.388$\pm$0.002 & 1.786$\pm$0.002 & 4.396$\pm$0.027 \TBstrut\\
1.85<P$_{rot}$<4.0 & 1.420 & 1.244$\pm$0.002 & 1.757$\pm$0.000 & 0.806$\pm$0.001 \TBstrut\\
4.0<P$_{rot}$<7.0  & 1.382 & 1.528$\pm$0.004 & 1.429$\pm$0.008 & 12.163$\pm$1.498 \TBstrut\\
7.0<P$_{rot}$<30.0 & 1.354 & 1.166$\pm$0.018 & 2.066$\pm$0.013 & 1.482$\pm$0.023 \TBstrut\\
30.0<P$_{rot}$ &1.300  &  &  &   \TBstrut\\
 \hline
\multicolumn{1}{l}{}&\multicolumn{4}{c}{0.3$\leq$<M$_\star$<0.45 M$_\odot$}\TBstrut\\
 \hline
P$_{rot}$<0.6      & 1.376 & 1.487$\pm$0.004 & 1.987$\pm$0.005 & 10.008$\pm$0.076 \TBstrut\\
0.6<P$_{rot}$<1.85 & 1.374 & 1.288$\pm$0.002 & 1.727$\pm$0.005 & 13.958$\pm$0.148 \TBstrut\\
1.85<P$_{rot}$<4.0 & 1.290 & 1.296$\pm$0.002 & 2.196$\pm$0.004 & 2.547$\pm$0.008 \TBstrut\\
4.0<P$_{rot}$<7.0  & 1.471 & 1.586$\pm$0.001 & 1.943$\pm$0.139 & 61.617$\pm$13.294 \TBstrut\\
7.0<P$_{rot}$<30.0  & 1.484 & 1.288$\pm$0.006 & 1.738$\pm$0.003 & 1.552$\pm$0.016 \TBstrut\\
30.0<P$_{rot}$ & 1.258 & 1.318$\pm$0.001 & 2.237$\pm$0.001 & 0.724$\pm$0.001 \TBstrut\\
 \hline
\multicolumn{1}{l}{}&\multicolumn{4}{c}{M$_\star\geq$0.45M$_\odot$} \TBstrut\\
 \hline
P$_{rot}$<0.6      & 1.326 & 1.171$\pm$0.004 & 1.645$\pm$0.002 & 3.067$\pm$0.022 \TBstrut\\
0.6<P$_{rot}$<1.85 & 1.224 & 1.286$\pm$0.002 & 1.761$\pm$0.004 & 6.822$\pm$0.056 \TBstrut\\
1.85<P$_{rot}$<4.0 & 1.139 & 1.138$\pm$0.001 & 2.051$\pm$0.001 & 1.490$\pm$0.001 \TBstrut\\
4.0<P$_{rot}$<7.0  & 1.492 & 1.378$\pm$0.001 & 1.697$\pm$0.001 & 2.358$\pm$0.011 \TBstrut\\
7.0<P$_{rot}$<30.0 & 1.060 & 1.099$\pm$0.001 & 2.356$\pm$0.001 & 0.916$\pm$0.000 \TBstrut\\
30.0<P$_{rot}$ & 1.500 & 1.537$\pm$0.014 & 2.330$\pm$0.026 & 0.604$\pm$0.010 \TBstrut\\
\end{tabularx}}

   \tablefoot{(1) These values were calculated with SimplexLSQFitter, which while effective for optimization, lacks a formal error estimation  and operates without derivatives.}

\end{table}
\begin{figure}
\sidecaption
\includegraphics[width=8cm]{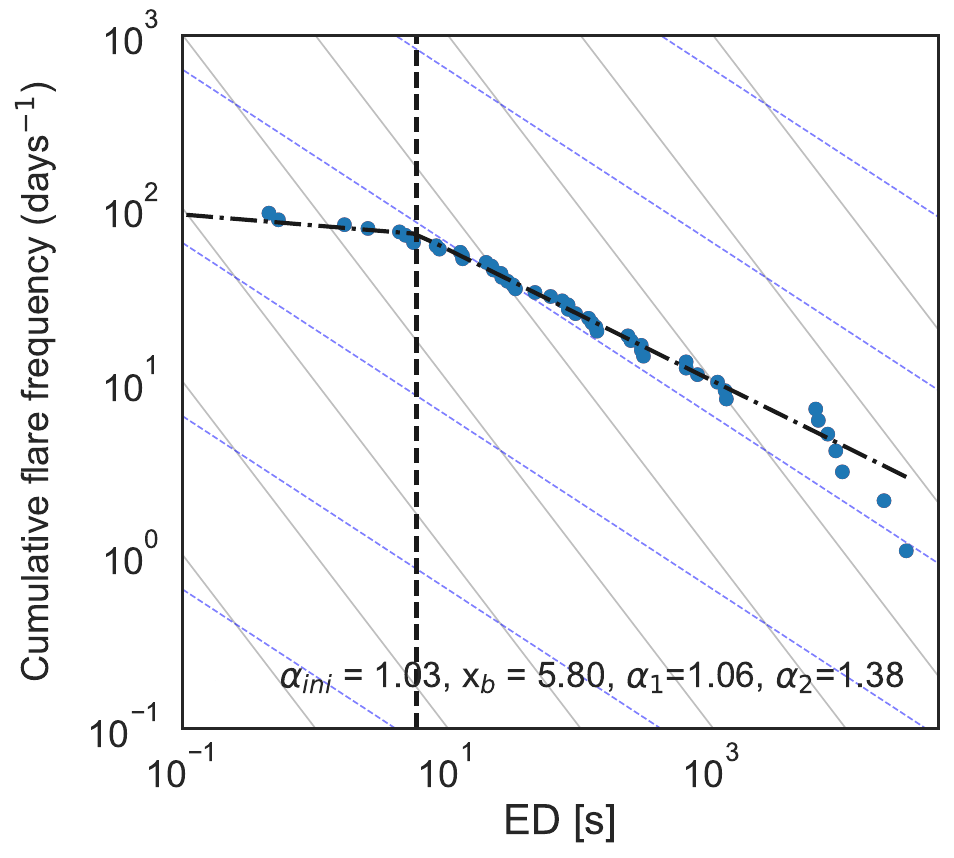}
\caption{Cumulative FFDs (scatter) in ED, and respective broken power law fits (dashed black lines) for the sample of FUV flares in young and field stars in this study sample. The y-axis shows the frequency distribution plotted against ED on the x-axis. The grey and blue dashed guides corresponding to power law coefficient $\alpha$=2.0 and $\alpha$=1.5, respectively, for a range of ED.}

\label{fig:45}
\end{figure}


\begin{figure*}\resizebox{\hsize}{!}{
   \includegraphics{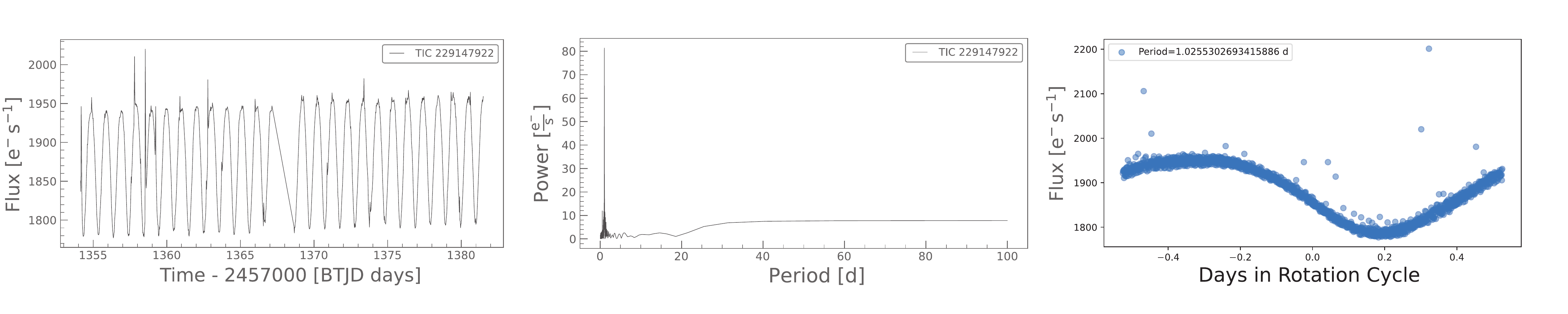}
    }
\caption{Example of a light curve (left), a periodogram (centre) and a folded light curve (right) for UCAC4 208-001676 in TESS Sector 1 with the fitted frequency of 1.03 days found by the Lomb-Scargle method. At the centre panel, the period at maximum power, found by the periodogram method, is clearly seen as a peak value. We analysed four sectors of TESS data for this star: 2, 3, 29, 30, 69 and the average period found to be 1.02 days.}
\label{fig:3}
\end{figure*}

\begin{figure*}\resizebox{\hsize}{!}{
   \centering
   \includegraphics{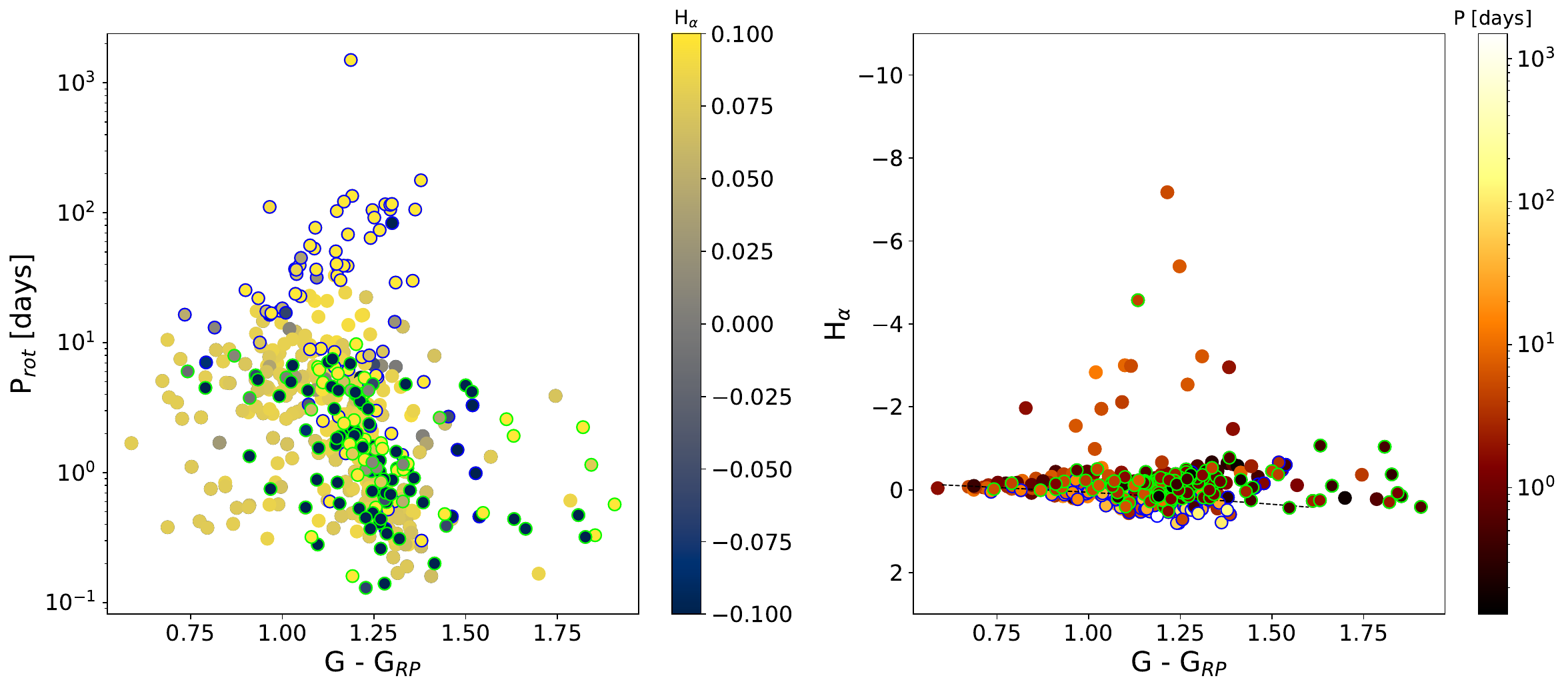}
    }
\caption{To compliment our analyses in discrepancies between EW H$\alpha$ measured by ESP-ELS in Gaia DR3 and the values reported in literature, here we plot the relationship between H$\alpha$ from Gaia DR3 including the model (see Eq.\ref{eq:11}) and logP$_{rot}$ across (G - G$_{RP}$)) colour from Gaia DR3.   Left: the rotation period distribution with (G - G$_{RP}$)) colour-coded by normalised H$\alpha$ equivalent width. Right: H$\alpha$ equivalent width against (G - G$_{RP}$), colour-coded by logP$_{rot}$. Stars with newly determined periods P$_{rot}$ are plotted as circles with lime-green edges, field stars in the sample are represented as circles with blue edges. The M-dwarf activity boundary plotted as dashed line follows \citet{2021AJ....161..277K}}
\label{fig:36}
\end{figure*}

\begin{figure*}\resizebox{\hsize}{!}{
   \centering
   \includegraphics{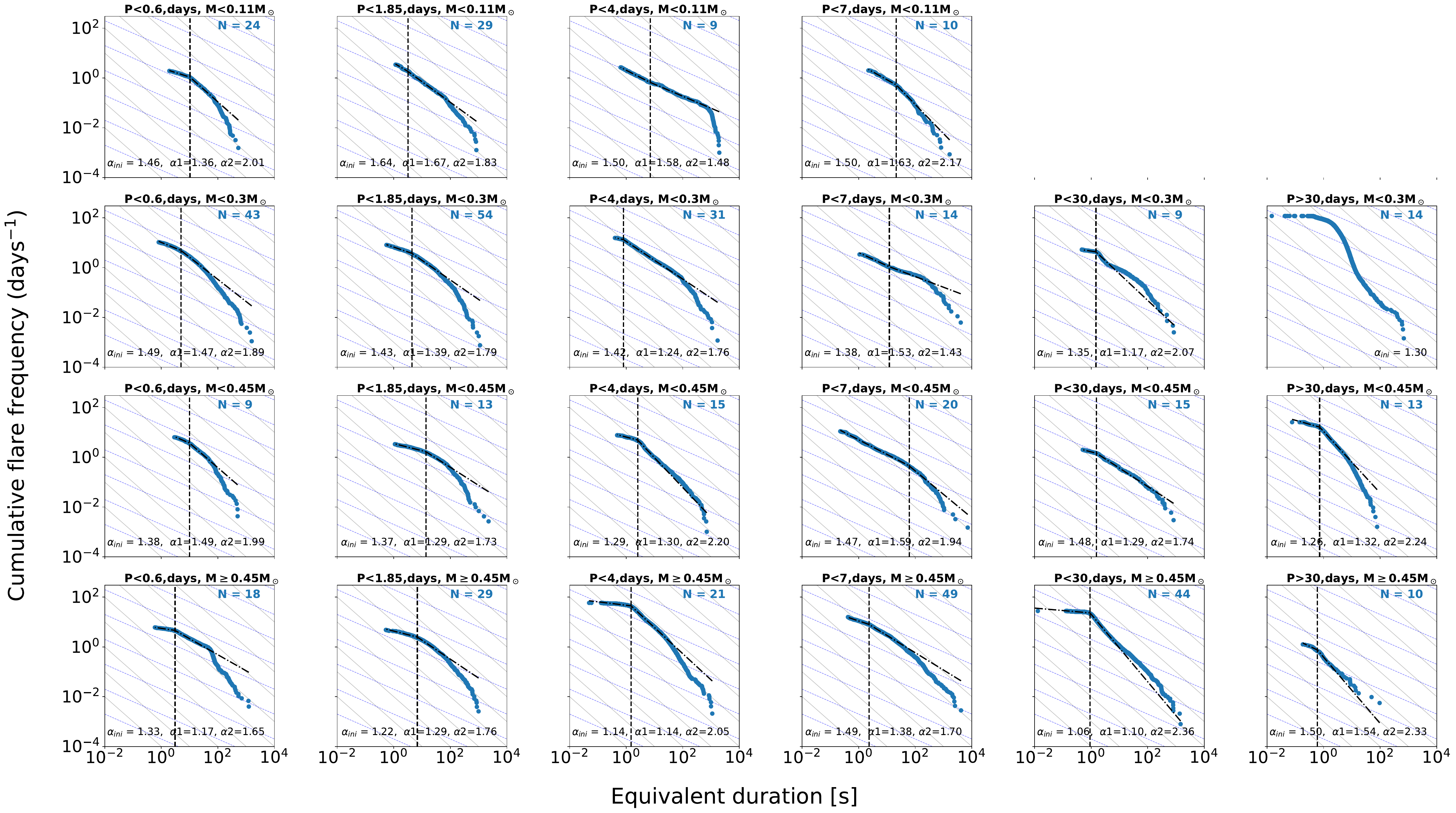}
    }
\caption{The cumulative distribution for the mass-period bins of all stars in this study sample. Blue dots represent the bin subsample FFD, which takes to account multiple stars by averaging each portion of the FFD by the number of stars that contribute to it. Initial guess for slope $\alpha_{1~ini}$ and intercept $\beta_{ini}$ from MMLE method as before were used to fit the broken power law was to the distribution resulting slopes $\alpha_1$, $\alpha_2$ (black dash-dotted lines) and break points indicated by black dashed lines, if the fit was successful. The grey and blue dashed guides corresponding to power law coefficient $\alpha$=2.0 and $\alpha$=1.5, respectively, are plotted for a range of ED $\in$[10$^{-2}$, 10$^{4}$] days. }
\label{fig:25}
\end{figure*}

\begin{figure*}\resizebox{\hsize}{!}{
   \centering
   \includegraphics{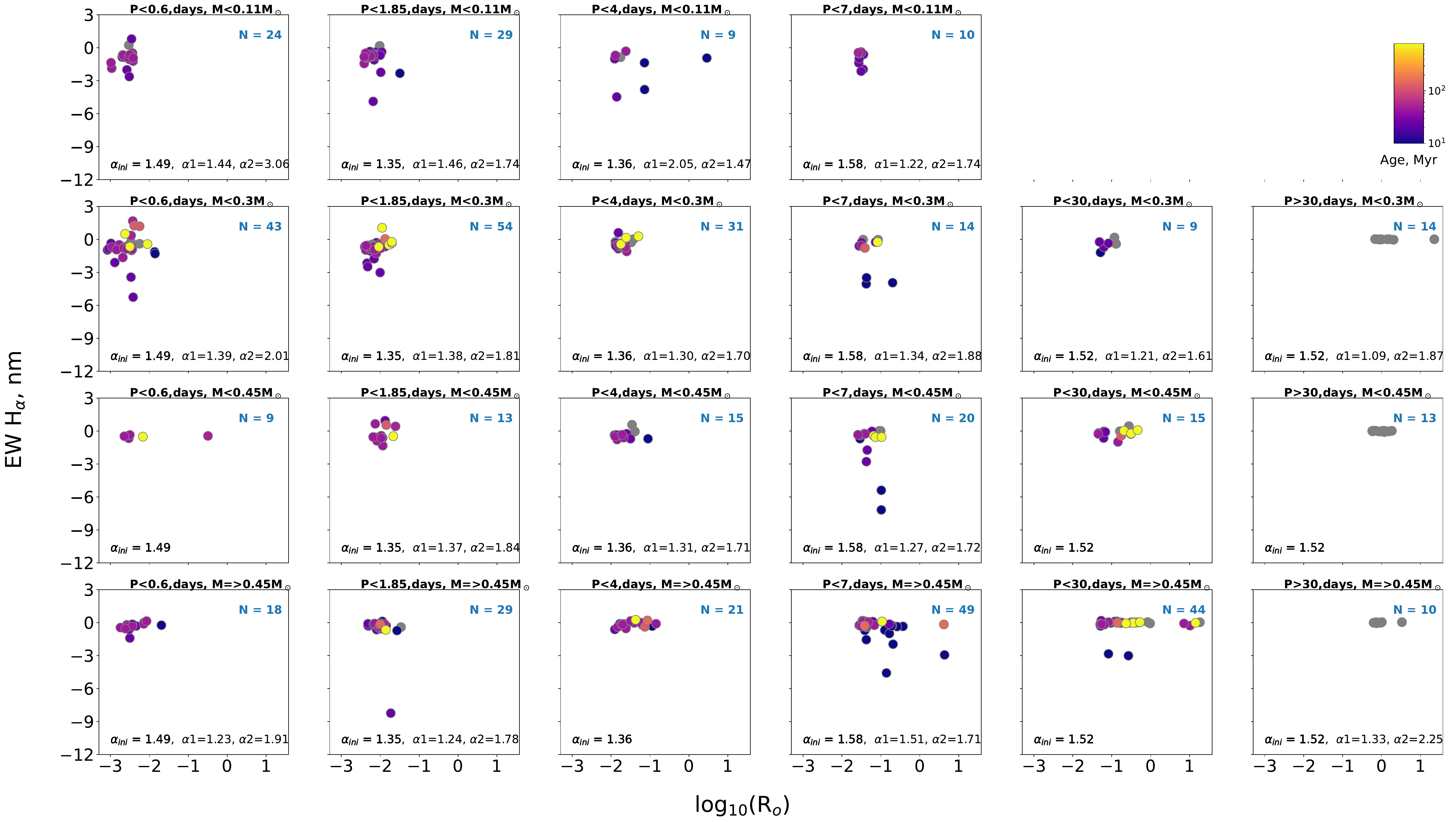}
    }
\caption{The relation between the Rossby number and EW H$\alpha$. The dots are colour-coded according to the age of the star. Initial guess for slope $\alpha_{ini}$ along with the broken power law slopes $\alpha_1$ and $\alpha_2$ for successful fits are stated for every mass-period bin.}
\label{fig:46}
\end{figure*}
\end{appendix}

 \end{document}